%% file: Revised_manuscript.tex
\documentclass[a4paper,doc,natbib]{apa6}

\usepackage[english]{babel}
\usepackage{appendix}
\usepackage[utf8]{inputenc}
\usepackage{latexsym,amsmath,amssymb,amsthm,amsfonts,graphicx,natbib}
\usepackage{bigints}  
\usepackage{xcolor} 
\usepackage{float}
\usepackage{multirow}
\usepackage{rotating}
\usepackage{mathtools}
\usepackage{hyperref}
\hypersetup{
	colorlinks=true,
	linkcolor=blue,
	filecolor=magenta,   
	citecolor={blue},   
	urlcolor=blue,
	pdftitle={Overleaf Example},
	pdfpagemode=FullScreen,
}

\include{Notations}
\graphicspath{{Images/}}

\newtheorem{lemma}{Lemma}
\title{Efficient Selection Between Hierarchical Cognitive Models: Cross-validation With Variational Bayes}
\shorttitle{Cross-validation with Variational Bayes}
\author{Viet-Hung Dao$^1$, David Gunawan$^2$, Minh-Ngoc Tran$^3$, Robert Kohn$^1$, Guy E. Hawkins$^4$, Scott D. Brown$^4$}

\affiliation{1: Australian School of Business, University of New South Wales, Sydney, Australia \\
 2: School of Mathematics and Applied Statistics, University of Wollongong \\
 3: Discipline of Business Analytics, University of Sydney Business School \\
 4: School of Psychology, University of Newcastle, Australia}
\date{\today}

\begin{document}

\maketitle

\begin{abstract}

Model comparison is the cornerstone of theoretical progress in psychological research. Common practice overwhelmingly relies on tools that evaluate competing models by balancing in-sample descriptive adequacy against model flexibility, with modern approaches advocating the use of marginal likelihood for hierarchical cognitive models. Cross-validation is another popular approach but its implementation remains out of reach for cognitive models evaluated in a Bayesian hierarchical framework, with the major hurdle being its prohibitive computational cost. To address this issue, we develop novel algorithms that make variational Bayes (VB) inference for hierarchical models feasible and computationally efficient for complex cognitive models of substantive theoretical interest. It is well known that VB produces good estimates of the first moments of the parameters, which gives good predictive densities estimates. We thus develop a novel VB algorithm with Bayesian prediction as a tool to perform model comparison by cross-validation, which we refer to as CVVB. In particular,  CVVB can be used as a model screening device that quickly identifies bad models. We demonstrate the utility of CVVB by revisiting a classic question in decision making research: what latent components of processing drive the ubiquitous speed-accuracy tradeoff? We demonstrate that CVVB strongly agrees with model comparison via marginal likelihood, yet achieves the outcome in much less time. Our approach brings cross-validation within reach of theoretically important psychological models,  making it feasible to compare much larger families of hierarchically specified cognitive models than has previously been possible. To enhance the applicability of the algorithm, we provide Matlab code together with a user manual so users can easily implement VB and/or CVVB for the models considered in this paper and their variants.

\end{abstract}
	
\vspace*{1cm}	

\section{Introduction}

Progress in psychological science can be made by choosing between competing theories: Does sleep deprivation cause attentional lapses? Does alcohol impair the speed of information processing or reduce cautiousness, or both? Does the forgetting curve follow a power or exponential function? When these theories are quantitative models that can be estimated from observed data (i.e., ``fitted''), the problem is known as model selection. Model selection continues to be a thorny problem for psychological researchers, even after many decades of progress \cite[e.g.,][]{Myung2000,RobertsPashler2000,navarro2019between,gronau2019limitations}. The key difficulty in model selection is balancing goodness of fit against model flexibility; that is,  balancing the degree to which each model accounts for the patterns observed in data against its ability to predict related future  data patterns. Model flexibility is often defined as the range of data patterns that a model can predict, which includes patterns that were observed as well as patterns that were not observed. Overly-flexible models are theoretically non-informative because they can ``predict'' almost any pattern that could be observed.

Many approaches were developed to tackle this problem. These include likelihood ratio tests, various information criteria, e.g., Akaike, Bayesian and Deviance Information Criteria; AIC, BIC, and DIC, respectively, minimum description length, and marginal likelihood, i.e., Bayes factors. Among these, cross-validation is the most popular \citep{EfronGong1983,Browne2000,vehtari2014waic}. A key strength of cross-validation is that it directly asks the question that scientists are often interested in: how well will this model predict new data? The simplest version of cross-validation divides observed data into two disjoint and approximately equal parts. The first, the  ``estimation'' subset, is used to estimate the model, while the second, the  ``validation'' subset, is held out. The procedure is  repeated with the second subset used to estimate the model and the first subset is used for  validation. The average of the validation performance measures, such as mean squared errors (MSE) is then used to compare different models.  The model is evaluated on its ability to predict the held-out data, treating them as new observations.

While cross-validation is widely agreed to be a desirable method for model selection, it is not used very widely in psychological science. A principal reason for this is its computational cost. Cross-validation is usually carried out repeatedly, using many different ways of splitting the observed data in the estimation and validation subsets; this is important in order to reduce sampling error associated with implementing the subsetting. Leave-one-out cross-validation (LOO-CV) leaves out
one data point at a time and uses the rest of the data to estimate the model. LOO-CV is closest to actual prediction but it is computationally extremely expensive. A more practical version is $K$-fold cross-validation ($K$-fold CV) in which the data is partitioned into $K$ folds (a common choice is $K=5$ or 10). It is implemented with one fold
left out as the validation subset and the model is estimated based on the other folds. This 
effectively requires estimating the model
on a ``new'' subset of estimation data $K$ times, which can be particularly time consuming in modern quantitative psychology, given the emphasis on using hierarchical Bayesian methods. Hierarchical models include random effects to account for the ubiquitous differences between human participants. With complex psychological theories, evaluating such a model in a Bayesian framework can take days of computer time, which makes repeated evaluation for cross-validation impractical. For example, one approach to understanding which elements of a particular psychological theory may be critical to explaining observed data is to enumerate a large family of model variants. These model variants are formed from all the combinations produced by
including or excluding different elements that have been hypothesized to be important, leading to dozens or even hundreds of models to consider. Using cross-validation to choose between such a large set of models is even less practical. As long as this approach to cross-validation remains out of reach, an unresolved issue is that model selection is subject to researcher bias: researchers select and then compare the subset of models they believe to be a priori most reasonable, since we cannot enumerate and feasibly compare all possible models.

We propose a solution to this problem, allowing cross-validation to be used very efficiently with complex psychological theories which include random effects. Our approach maintains the hierarchical Bayesian structure of the models, but employs variational Bayes to increase the computational efficiency so greatly that cross-validation becomes practical. Variational Bayes (VB; also known as variational approximation, VA) methods provide an alternative to more widely-used methods based on Markov chain Monte-Carlo (MCMC). VB methods have become increasingly prominent for approximate Bayesian inference in a wide range of challenging statistical models \citep[for reviews, see, e.g.,][]{blei2017variational,ormerod2010explaining}. With VB, the problem of estimating the posterior distribution is re-formulated as an optimization problem. The (usually very complex) posterior distribution of interest is approximated by a simpler, more tractable distribution that is selected to balance accuracy and computational cost. The parameters of the approximating distribution are then identified by an optimization which minimizes the Kullback-Leibler distance between the approximating distribution and the posterior distribution. With  careful choices of the approximating distribution and optimization algorithm, VB methods can produce results 10 or 100 times faster than exact methods such as MCMC. However, unlike MCMC, variational methods are approximate.

Despite their strengths, VB methods are still not widely used in psychological research \citep[see, however,] []{galdo2019variational}. One reason is that VB methods have certain limitations which make drawing model-based inferences difficult. The quality of the approximation is not always well-known; the methods have a tendency to underestimate the variability of the posterior distribution, and this can be problematic for parameter inference such as setting credible intervals as well as model selection based on the marginal likelihood. A key insight underpinning our work is that VB methods are quite accurate at estimating the posterior means
\citep[see, for example, the discussion in][]{blei2017variational}, even though
they often underestimate the posterior  variances.
This is a crucial distinction for application to cross-validation. In cross-validation, the  performance
of a model is evaluated on how well it predicts held-out data, and here the role of the posterior variances is of second order importance at most.
We show in a simulation study that the predictive densities estimated by MCMC and VB are very close when  VB underestimates some of the posterior variance of the model parameters.

Following the above discussion, and building on recent work using VB methods in cognitive science by \citet{galdo2019variational}, we propose   combining
cross-validation and VB, which we call ``CVVB''. By employing modern VB methods, we show that CVVB can handle complex psychological theories with random effects in hierarchical Bayesian frameworks. Even with such challenging models, CVVB is sufficiently fast to make it practical to use when searching a large number of competing models, as described above;  an example below illustrates selection amongst 256 competing models. Alternatively, for those who prefer using exact Bayesian model selection approaches, such as marginal likelihood, CVVB may be used as an efficient model screening tool. That is, when given a very large set of models to evaluate,  CVVB can be used to screen out the poorest-performing models. This reduces the number of candidate models to a manageable size, and slower exact Bayesian methods (e.g.,  marginal likelihood) can then be used on the remaining models. 

The article first outlines the VB method and then develops the novel VB algorithms.
These algorithms are presented in a general way to make their implementation apparent for a range of psychological models. The performance of the novel VB algorithms is demonstrated
in a cognitive model for decision-making. Following this,  the CVVB procedure is developed
through a detailed example of the model selection approach, continuing the analysis of the cognitive model for decision-making. The example enables us to revisit a theoretical question about the speed-accuracy tradeoff in decision-making \citep{rae2014hare,StarnsRMcK2012,Lee2008};
the question was previously addressed by statistical model selection methods with the shortcomings described above. Using CVVB, we are able to address the question of key scientific interest using more advanced model selection methods. Matlab code for the methods can be found at \href{https://github.com/Henry-Dao/CVVB}{https://github.com/Henry-Dao/CVVB}. We also provide a detailed user manual that explains the inputs and outputs of the algorithms, 
how to run the examples in the paper, and
how to modify the code to apply it to new models. 

\section{Variational Bayes}\label{sec: Intro VB}

This section introduces the basic ideas behind VB methods. We focus on the ``fixed form'' method, also known as stochastic VB, which is currently widely used in the machine learning and statistics literatures. We then introduce particular applications of the method, which we will focus on in this article. These methods are particularly well-suited to applications in psychology, where almost all models include random effects (for participants) and have correlated parameters (due to the overlapping and inter-dependent nature of the underlying psychological constructs being modelled).

Bayesian model selection involves choosing between competing models (including priors). The basic model is defined by its likelihood function $\like$, which gives the probability density for observing data $y$ given parameters $\btheta = (\theta_1,\dots,\theta_p)$. In the Bayesian approach, the model parameters are governed by a prior distribution $\prior$ which encodes existing knowledge about plausible values for those parameters. The goal of inference is to estimate the posterior distribution $\post$, which expresses the plausibility of different parameter values, given the data. Closed-form solutions for the posterior distribution are rarely available, so Bayesian analysis requires methods for approximating the posterior distribution. Markov chain Monte Carlo (MCMC) produces simulation consistent Bayesian inference, i.e., we obtain exact answers as the number of MCMC iterates increases. A key disadvantage of MCMC methods for psychological models is that they can be very inefficient computationally when the posterior distribution is high-dimensional, i.e., the model has many parameters, or when the model's parameters are strongly correlated \citep{turner2013method}.

Variational Bayes (VB) is an approximate method to estimate the posterior. It is based on optimization: an easy-to-use distribution is chosen to approximate the posterior distribution, and then parameters for the approximating distribution are found by optimizing the fit to the posterior distribution. Let $\qvb$ denote the approximating distribution for $\btheta$ which has parameters $\blamb$ called the variational parameters. The best choice for these parameters is identified by minimizing the Kullback-Leibler (KL) divergence between the approximating distribution and the posterior distribution:
$$KL(\qvb||\post) := E_{\qvb}\left[ \log\dfrac{\qvb}{\post}\right]\,\footnote{The notation $E_f$ denotes the expectation with respect to distribution $f$. }.$$
The KL divergence has the property that $KL(\qvb||\post) \geq 0$ with  equality if and only if
$\qvb = \post$.  Since, 
\begin{align*}
	0\leq KL(\qvb||\post) &= E_{\qvb}\left[ \log\qvb - \log\post \right] \\
	&= E_{\qvb}\left[ \log\qvb - \log\joint + \log p(\by) \right]\\
	&=E_{\qvb}\left[ \log\qvb - \log\joint \right] + \log p(\by);
\end{align*}
therefore, 
$$ \log p(\by)\geq \lb:= E_{\qvb}\left[\log\joint   - \log\qvb\right]. $$

Hence, minimizing the KL divergence between $\qvb$ and $\post$ is equivalent to maximizing $\lb$, which is called the lower bound. This allows optimization of the fit between the approximating and posterior distributions to proceed by searching on parameters $\blamb$ to maximize the quantity $\lb$. The search can be computationally difficult, if the approximating distribution has many parameters or is chosen poorly. Our approach relies on recent developments in the statistical literature to simplify the optimization. We apply stochastic-gradient search methods \citep{robbins1951stochastic}, and improve their precision using the reparameterization ``trick'' of \cite{kingma2013auto} and \citet{rezende2014stochastic}. We further simplify the problem by reducing the dimension of $\blamb$, using a factor structure for some of its parameters. Finally, we automate the problem of identifying separate step sizes for elements of the vector $\blamb$ using the adaptive learning and stopping rule developed by \cite{zeiler2012adadelta}. Appendix \ref{appendix:VB-details} gives the details. 

\subsection{Gaussian Variational Bayes with a Factor Covariance Structure}

Gaussian VB is the most common VB approach; here the variational distribution  $\qvb = \mvntheta$ is Gaussian\footnote{$\mvn$ denotes a $p$-dimensional normal distribution with mean $\bmu$ and covariance matrix $\bSig$; $\mvntheta$ denotes the corresponding multivariate normal density with argument $\btheta$.}. Gaussian VB is often motivated by the observation that the posterior can be well approximated by a normal distribution under general conditions, when there are sufficient data \citep{bernardo2009bayesian}. For a Gaussian approximating distribution, the dimension of $\blamb$ is $p+p(p+1)/2$. This means that the dimension of the parameters to be searched over in the approximation step increases quadratically with the number of model parameters -- due to all the covariance elements in the matrix $\bSig$. One way to simplify the optimization problem is to set $\bSig$ to a diagonal matrix, but this is unsatisfactory for psychological models because it makes the very restrictive assumption of posterior independence between the components
\citep[as in][]{turner2013method}.

Following \citet{ong2018gaussian}, we make the  covariance matrix parsimonious by using a standard factor structure; i.e., we assume that
$\bSig = BB^\top+D^2$, where $B$ is a $p\times r$ matrix and $D$ is a diagonal matrix with diagonal elements $d=(d_1,\dots,d_p)$. By choosing the number of factors $r \ll p$,   the factor approximation is simpler and the VB optimization is more tractable. The approximating distribution is normal, with mean $\mu$ and variance matrix $\bSig $, which  means that the size of the search problem is much smaller; the vector to be searched over is $\blamb = (\bmu^\top,\text{vec}(B)^\top,d^\top)^\top$\footnote{$\vect(A)$ is a column vector obtained by stacking the columns of matrix $A$ under each other from left to right.}. Approximating the posterior distribution by searching over $\blamb$ is made even more efficient  by applying the reparameterization trick to reduce the variance of the gradient estimate of the lower bound, leading to fast and accurate approximations of the gradient during the search (see Appendix \ref{appendix:VB-details}).

\section{Variational Bayes for Psychological Models with Random Effects\label{sec: VB for LBA}}

This section develops the Gaussian VB method presented in the previous section for Bayesian inference with hierarchical psychological models. In a hierarchical model, participants are allowed to have different values for one or more of the model parameters and such parameters are called random effects. These random effects capture the important psychological differences between participants, and avoid many of the problems associated with averaging across people. We make the model estimation more tractable by assuming that the random effects follow some group-level distribution, rather than being independent across people. Here, we assume that the distribution of random effects in the population is multivariate normal, possibly after an appropriate parameter transformation.

The application of simple Gaussian VB to a generic cognitive model that is defined by some arbitrary density function is first illustrated. The approximation is then improved
by exploiting the structure of hierarchical cognitive models.

Suppose there are $J$ participants who all perform a cognitive task, with each subject completing multiple trials; on each trial, a stimulus is presented and the subject produces a response. For participant $j$, the observed response on trial $i$ is denoted $y_{j i}$, with $y_{j i}$  generated by $p(y_i|\balph_j)$, the density function of the observations according to the cognitive model,
where $\balph_j = (\alpha_{j1},\dots,\alpha_{j D_\alpha})$ is the vector of $D_\alpha$ parameters. The $n_j$ responses from participant $j$ are denoted $\by_j=(y_{j1},\dots,y_{j n_j})$ and the collection of responses from the sample of $J$ participants is $\by=(\by_1,\dots,\by_J)$. With the usual assumptions of independence between trials, the conditional density of all the observations is
\begin{equation}
	p(\by|\balph) = \prod\limits_{j=1}^J\prod\limits_{i=1}^{n_j} p(y_{ji}|\balph_j).
	\label{eq:p-y-given-alpha-J}
\end{equation}
We assume the elements of $\balph_j$ have support on the real line (possibly after transformation).
This assumption makes it possible to assume a multivariate normal distribution for the group-level distribution of the random effects. The full model for the data is,
\begin{enumerate}
	\item Conditional density: 	$y_{ji}|\balph_j \stackrel{i.i.d.}{\sim}  p(y_{ji}|\balph_j)$ for $j = 1,\dots,J; \, i = 1,\dots,n_j$.
	\item A multivariate normal distribution for the random effects
	\begin{align}
		\balph_j|\bmualph,\bSigalph \stackrel{i.i.d.}{\sim} N(\bmualph,\bSigalph ).\label{eq:prior-random-effects}
	\end{align}
	\item Priors for model parameters: We follow \citet{gunawan2020new} and use a normal prior for $\bmualph$ and the marginally non-informative prior for $\bSigalph$ suggested by \cite{huang2013simple}:
	\begin{align}
		\begin{array}{l}
			\bmualph \sim N(\boldsymbol{\boldsymbol{0}},\boldsymbol{I}),\\
			\bSigalph|a_1,\dots,a_{\Da} \sim \textrm{IW} \left(\Da +1,\bPsi\right),\bPsi=4\textrm{diag}\left(\dfrac{1}{a_1},\dots,\dfrac{1}{a_{\Da}}\right),\\
			a_1,\dots,a_{\Da} \sim \textrm{IG} \left( \dfrac{1}{2},1\right).
		\end{array}\label{eq:prior-model-parameters}
	\end{align}
\end{enumerate}
The notation $\textrm{IW}(\nu,A)$ denotes an inverse Wishart distribution with degrees of freedom $\nu$ and scale matrix $A$ and $\textrm{IG}(1/2,1)$ denotes an inverse Gamma distribution with scale parameter $1/2$ and shape 1.

\subsection{Gaussian Variational Bayes}

The parameter vector of the psychological model, $\btheta$, includes random effects for every subject ($\balph_{1:J})$, the group-level mean ($\bmualph$) and variance ($\bSigalph$) parameters, as well as the hyperparameters $\ba = (a_1,\dots,a_{\Da})$ of the prior. The random effects ($\balph$) and the group-level means ($\bmualph$) have support on the real line, but the covariance parameters ($\bSigalph$) are restricted to form a positive definite covariance matrix, and the hyperparameters $\ba$ are strictly positive. These constraints make it unreasonable to approximate the posterior distribution by a Gaussian distribution. To obtain a useful Gaussian variational approximation, we transform the parameters, where necessary, so that all the elements now have support on the full real line. Let $\bSigalph = \bC\bC^\top$ be the Cholesky decomposition of the group-level covariance matrix, with $\bC$ a lower triangular matrix with positive elements on the diagonal. We can therefore reparametrize $\bSigalph$ by an unconstrained vector lying on the real line consisting of the strict lower triangle of $\bC$ and the logarithms of the diagonal elements of $\bC$. We similarly log-transform the hyperparameters $\ba$. The working parameters are $\tilde{\btheta} = (\balph_1^\top,\dots,\balph_J^\top,\bmualph^\top,\vech(\bCstar)^\top,\log(a_1),\dots,\log(a_{\Da}))^\top$,\footnote{$\vech(A_{n\times n})$ is the $(n(n+1)/2)\times 1$ column vector obtained from $\vect(A)$ by omitting all the upper triangular elements of $A$.} with $\bCstar$ indicating the lower triangle of matrix $\bC$. Appendix \ref{appendix:gradients-GVB-Hierarchical-LBA} gives the technical details.

\subsection{Hybrid Gaussian Variational Bayes}

We now develop a novel extension to Gaussian VB for hierarchical models with random effects, which exploits the structure of the posterior distribution. In the hierarchical models we consider, the posterior distribution can be factored as
$$p(\balph_{1:J},\bmualph,\bSigalph,\ba|\by) = p(\balph_{1:J},\bmualph,\ba|\by)p(\bSigalph|\balph_{1:J},\bmualph,\ba,\by).$$
It is not difficult to show that the conditional density $p(\bSigalph|\balph_{1:J},\bmualph,\ba,\by)$ is the density of $\textrm{IW}(\bSigalph|\nu,\bPsi')$
with $\nu = 2\Da + J + 1$ and $\bPsi' = \sum_{j=1}^J (\balph_j - \bmualph)(\balph_j - \bmualph)^\top + 4\diag\left( {1}/{a_1},\dots,{1}/{a_{\Da}}\right)$(Appendix \ref{appendix:gradients-hybrid-GVB}, Lemma \ref{lemma: full-conditional-density}).

This suggests that it is only necessary to approximate the joint posterior of the random effects vectors ($\balph_{1:J}$), the group-level mean parameters ($\bmualph$), and the hyperparameters ($\ba$). That is, we use a VB approximating distribution, $\qvb$, of the form
$$q_{\blamb}(\balph_{1:J},\bmualph,\ba,\bSigalph) = q_{\blamb}(\balph_{1:J},\bmualph,\ba) \textrm{IW}(\bSigalph|\nu,\bPsi').$$
This hybrid variational distribution takes into account the posterior dependence between $\bSigalph$ and the other parameters, which allows for a more accurate approximation to the posterior.
The set of parameters is now $\tilde{\btheta}=(\balph_{1:J},\bmualph,\log \ba,\bSigalph)$ and the data-parameter joint density becomes
\begin{align*}
	p(\by,\balph_{1:J},\bmualph,\bSigalph,\log \ba) =& \prod\limits_{j=1}^J f(\by_j|\balph_j)N(\balph_j|\bmualph,\bSigalph)N(\bmualph|\boldsymbol{\boldsymbol{0}},\boldsymbol{I})\textrm{IW}(\bSigalph|\nu,\bPsi)\\
	&\times \prod\limits_{d=1}^{\Da}\textrm{IG}(a_d|1/2,1)\left| J_{a_d\rightarrow\log a_d}\right|,
\end{align*}
where $J_{a_d\rightarrow\log a_d} = a_d$ is the Jacobian of the transformation.

If  the parameters are separated as $\btheta_1 = (\balph_{1:J},\bmualph,\log \ba)$ and $\btheta_2=\bSigalph$ and   $q_{\blamb}(\btheta_1)$ is parameterized by a Gaussian density that assumes a reduced factor structure for its covariance matrix, then the variational distribution has the  parametric form
$$q_{\blamb}(\btheta_1,\bSigalph) = N(\btheta_1|\bmu,BB^\top + D^2) \textrm{IW}(\bSigalph|\nu,\bPsi'),$$
with the variational parameters $\blamb = (\bmu,B,d)$ (recall $D$ is a diagonal matrix with the diagonal vector $d$).
We refer to this approach as \emph{Hybrid Gaussian VB}.
We can write $\btheta_1 = u(\beps;\blamb):= \bmu + B\beps_1 + d\odot\beps_2$, with $\beps=(\beps_1^\top,\beps_2^\top)^\top\sim N(\boldsymbol{0},\boldsymbol{I})$. Using the reparameterization trick, the lower bound can be written as
$$\lb = E_{(\beps,\btheta_2)}\left[ \log p(\by,u(\beps;\blamb),\btheta_2) - \log q_{\blamb}(u(\beps;\blamb)) - \log p(\btheta_2|u(\beps;\blamb),\by) \right].$$

The idea of hybrid VB is also explored recently by \cite{Loaiza-Maya2020};
however, they do not include the term
$\nabla_{\btheta_1}\log p(\btheta_2|\btheta_1,\by)$ in their calculation of the lower bound gradient. Appendix \ref{appendix:VB-details} gives details for the gradient function of this lower bound, including efficient estimation methods based on the work of \cite{Loaiza-Maya2020}.

\section{CVVB: Model Selection by Variational Bayes with $K$-Fold Cross-validation}

The aim of cross-validation (CV) is to assess how well a model will predict out of sample. There are several versions of CV \citep{arlot2010survey}. The popular $K-$fold CV divides the data into $K$ approximately equal parts called \lq folds\rq. The model is first estimated using folds 2 to $K$, (the ``estimation data'') and then the estimated model is used to predict the data in the first fold (the ``validation data''). This is then repeated with folds 2 to $K$ successively left out of the estimation and used for model validation. CV can be computationally expensive as the process must be repeated many times, holding out a different fold each time.

This section describes a strategy for speeding up $K$-fold cross-validation based on VB, and refer to the method as cross-validation variational Bayes (CVVB). Our approach is based on two key observations. First, VB is very fast and is also good for prediction \citep{blei2017variational}. Second, when the data are randomly split into folds of similar sizes, the VB approximations should not differ much across the data folds. Because of this, we can initialize the VB search algorithm for every fold after the first one using the results of the first estimation. Good initialization is important in VB optimization and helps to significantly speed up the convergence.

CVVB can be used as a model selection method by choosing the best model based on predictive performance in the held-out data. Alternatively, for those who prefer exact Bayesian methods, CVVB may be used as a model screening tool. That is, when given a very large set of models to evaluate, one can use CVVB to efficiently screen out the poorest-performing models. This reduces the set of candidate models to a manageable size, and it is then possible to use slower exact Bayesian methods (such as the marginal likelihood) on the remaining models.

An important choice in $K$-fold CV is the choice of
loss function for the validation fold. In principle, almost any statistic which summarizes the discrepancy between the model's predictions and the held-out data is adequate. In Bayesian statistics, predictive performance is most commonly measured by the expected log predictive density (ELPD) \citep{gelman2013bayesian}:
$$\textrm{ELPD} := \int \log p(\tilde{\by}|\by) p^*(\tilde{\by}) d\tilde{\by};$$
$p^*(\tilde{\by})$ is the unknown true distribution of future observations $\tilde \by$, and $p(\tilde{\by}|\by)$ is the posterior predictive density. This is the density of the future observations, integrated over the posterior distribution of the parameters:
\[p(\tilde{\by}|\by) = \int p(\tilde{\by}|\btheta)\post d\btheta. \]

It is straightforward to estimate ELPD by CV. The data are partitioned into $K$ folds of similar sizes $\by^{(k)},k=1,\dots,K$ (a typical choice of $K$ is 5 or 10). Let $\by^{(-k)}$ be the data after fold $k$ is left out. For random effect models, we partition the data in the subject level, i.e., the data from each subject is randomly split into $K$ disjoint subsets, hence $\by^{(k)}=(\by_{1}^{(k)},\dots,\by_{J}^{(k)})$ consists of observations from all subjects for fold $k$ (appendix E gives the details of CVVB applied to random effect models). The $K$-fold cross-validation estimate for ELPD is
\[ \elpdcv := \dfrac{1}{K}\sum\limits_{k=1}^K \log  p(\by^{(k)}|\by^{(-k)}).  \]

The term $p(\by^{(k)}|\by^{(-k)})$ is the posterior predictive density for the $k^{th}$ fold, and represents the log score when the data in that fold are treated as unseen, and predicted using the posterior distribution estimated from the other folds. Using VB methods, this posterior predictive density can be estimated by drawing $S$ samples from the variational distribution as
\begin{align*}
	p(\by^{(k)}|\by^{(-k)})&=\int p(\by^{(k)}|\btheta)p(\btheta|\by^{(-k)})d\btheta\\
	&\approx \int p(\by^{(k)}|\btheta)q_{\lambda^{(k)}}(\btheta)d\btheta\\
	&\approx \dfrac{1}{S}\sum\limits_{s=1}^S p(\by^{(k)}|\btheta^{(s)}),\textrm{ with } \btheta^{(s)}\sim q_{\lambda^{(k)}}(\btheta),\, s = 1,\dots,S.
\end{align*}
Here, $q_{\lambda^{(k)}}(\btheta)$ is the VB posterior approximation for the leave-$k$th-fold-out posterior $p(\btheta|\by^{(-k)})$. By replacing the posterior predictive density $p(\by^{(k)}|\by^{(-k)})$ with the VB approximation,  the $K$-fold CVVB estimate for ELPD is obtained as
\[ \elpdcvvb := \dfrac{1}{K}\sum\limits_{k=1}^K \log \left(\dfrac{1}{S}\sum\limits_{s=1}^S p(\by^{(k)}|\btheta^{(s)}) \right).  \]


Although it is necessary to run the VB algorithm $K$ times for $K$-fold CV, the
warm-up initialization strategy discussed above means that the time taken to run all $K$ repetitions is almost the same as running VB once  on the full data set. Using the samples from the VB approximating distribution ($q_{\lambda^{(k)}}(\btheta)$) rather than from the exact posterior ($p(\btheta|\by^{(-k)})$) means that we only obtain approximate inference. However, this loss is offset by a very large gain in computational efficiency, making  the CVVB approach very attractive for quickly screening a large set of competing models.

\section{An Illustrative Application of Variational Bayes: Decision-Making by Evidence Accumulation}

We now apply the novel VB methods to an evidence accumulation model (EAM) for decision making. EAMs propose that decisions between competing alternative outcomes are made by accumulating evidence in favour of each possible response. The accumulation continues until a pre-defined threshold level of evidence is exceeded, after which the response corresponding to the winning accumulator is executed. While all EAMs share this basic structure, they differ in the specific details of the accumulation process and threshold setting. EAMs have been used to address important theoretical and applied questions in psychology \citep[for reviews, see][]{donkin2018response,ratcliff2016diffusion}. For example, EAMs helped to resolve theoretical debates about the mechanisms which underpin the cognitive slowdown observed during healthy ageing. It has long been known that older adults respond more slowly in many cognitive tasks than younger adults. For many decades, age-related slowing was attributed to a decrease in the rate of information processing \citep[the famous ``generalized slowdown'' hypothesis;][]{salthouse1996processing}. By applying EAMs to the data of older and younger adults, it was observed that a large proportion of the age-related slowdown effect was caused by increased caution rather than a decreased rate of processing \citep{ratcliff2004comparison,ThaparEtAl2003,forstmann2011speed,starns2010effects}. This kind of result typifies the benefit of using cognitive models to address applied questions, sometimes known as ``cognitive psychometrics'' \citep{Batchelderinpress}. Important psychological insights are supported by choosing between competing theories, which are represented by different model variants; e.g., comparing an EAM in which processing rate differs between younger and older groups vs. an EAM in which caution differs.

We focus on the linear ballistic accumulator \citep[LBA;][]{brown2008simplest}, which is simpler than many other EAMs in that it assumes no competition between alternatives \citep{brown2005ballistic}, no passive decay of evidence \citep{usher2001time} and no within-trial variability \citep{ratcliff1978theory,stone1960models}. This simplicity permits closed-form expressions for the likelihood function for the model parameters, which supports advanced statistical techniques including Bayesian methods based on MCMC and particle algorithms \citep{turner2013method,gunawan2020new,tran2020robustly,wall2020identifying}.

Most modern applications of the LBA model include a hierarchical random effects structure for individual differences. Bayesian methods for inference with the hierarchical LBA were first developed by \cite{turner2013method}. Recent developments have increased the efficiency of these exact methods, and extended them to allow for correlation between random effects \citep{gunawan2020new}. Even though these newer MCMC methods are more efficient than earlier methods, the computation time can still be quite costly. For example, for an experiment with 100 subjects each of whom contributes 1,000 decisions it can take several hours to estimate the model on a high-performance computer. This computational cost is one of the primary motivations for exploring VB methods.

We use the VB methods developed above to explore LBA models of decision-making in three data sets, as well as in a simulation study. We then demonstrate that addressing model selection among a large class of competing models is both feasible and practical with the CVVB approach. The CVVB approach is then used to address, more comprehensively than previous analyses, a debate about the effects of caution vs. urgency on decision-making \citep{rae2014hare}.

\subsection{The LBA Model of Decision Making}

\begin{figure}[]
	\begin{center}
		\includegraphics[scale= 0.6]{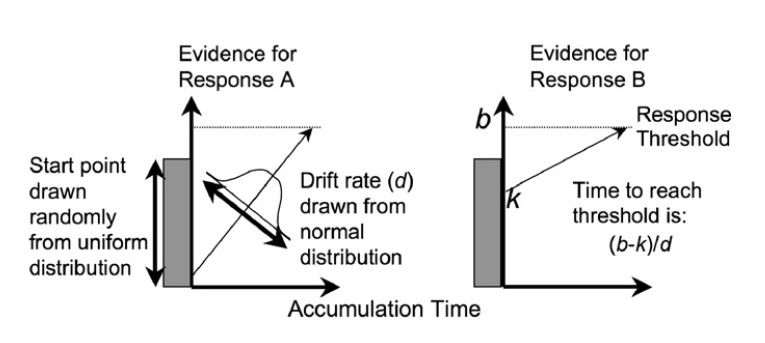}\\
	\end{center}
	\caption{An illustration of the LBA model for a binary choice with two evidence accumulators, one for ``Response A'' (left panel) and one for ``Response B'' (right panel). Evidence accumulates for each response until one reaches a threshold ($b$). The speed of evidence accumulation (drift rate $d$) and starting points ($k$) are random from decision to decision and between accumulators. }\label{fig: LBA_picture}
\end{figure}

The LBA model \citep{brown2008simplest} represents a choice between several alternatives as a race between different evidence accumulators, one for each response (see Figure \ref{fig: LBA_picture}); however, see \citet{van2019accumulating} for more flexible extensions. Each evidence accumulator begins the decision trial with a starting amount of evidence $k$ that increases at a speed given by the ``drift rate'' $d$. Accumulation continues until a response threshold $b$ is reached. The first accumulator to reach the threshold determines the response, and the time taken to reach the threshold is the response time (RT), plus some extra constant time for non-decision processes, $\tau$.

To explain the observed variability in the data, the model assumes that the starting points for evidence accumulators are random values drawn from a uniform distribution on the interval $[0,A]$, and the drift rates are drawn from normal distributions with means $v_1, v_2, \dots$ for the different response accumulators. It is usual to assume a common standard deviation $s$ for all accumulators \citep[but see also][]{DonkinPBR2009}. All random values are drawn independently for each accumulator, and are independent across decision trials. With these assumptions, \cite{brown2008simplest} and \cite{terry2015generalising} derive expressions for the distribution of the time to reach threshold, which we denote by $F_c$ and $f_c$, for the cumulative distribution function and probability density function, respectively. The joint density over response time $RT=t$ and response choice $RE=c$ is
$$ \textrm{LBA} (c,t|b,A,\bv,s,\tau) = f_c(t)\times \prod\limits_{k\neq c}(1-F_k(t)),$$
with $\bv = (v_1, v_2, \dots)$. Note that it is also possible to have parameters other than $v$ change between accumulators. For example, strategic decision biases may be represented by allowing different response thresholds ($b$) between accumulators. In these cases, the expression above generalizes in the obvious way, e.g., replacing the scalar parameter $b$ with a vector $\bb$.

The observed data from a single decision is represented by the vector of response time and choice, which we denote $y^\top_i = (RE_i,RT_i)$. If a participant provides a sequence of $n$ decisions, the vector of all data for the participant is denoted by $\by^\top = (y_1^\top,\dots,y_n^\top)$. Assuming independence across decision trials, the density for the data set is given by
$$ p(\by|b,A,\bv,s,\tau) = \prod\limits_{i=1}^n \textrm{LBA}(y_i|b,A,\bv,s,\tau), $$
For VB with the LBA model, this term replaces the generic model $p(\by|\balph)$.

\subsection{Hierarchical LBA Models}

We illustrate the generalization of the LBA model of how one person makes decisions to how a group of people make decisions with an example typical of the literature. \cite{forstmann2008striatum} collected data from 19 participants who performed a simple perceptual decision-making task. The participants were asked to decide, repeatedly, whether a cloud of semi-randomly moving dots appeared to move to the left or to the right. In addition, each participant was asked on some trials to respond very urgently, on other trials to respond very carefully, and on others to respond neutrally. These three speed-accuracy tradeoff conditions were of primary interest in the \citeauthor{forstmann2008striatum} analysis.

To capture the differences between the subjects, as well as the differences between the three conditions, \citet{gunawan2020new} proposed a hierarchical LBA model with three different threshold parameters $b^{(a)},b^{(n)}$ and $b^{(s)}$ for accuracy, neutral and speed conditions, respectively. They also proposed two parameters for the means of the drift rate distributions: one for drift rates in the accumulator corresponding to the correct response on each trial ($v_c$) and the other for the error response ($v_e$). Gunawan et al. assumed that the standard deviation of the drift rate distribution was always $s=1$. With these assumptions, each subject $j$ has the vector of random effects
$$\bz_j = (b^{(a)}_j,b^{(n)}_j,b^{(s)}_j,A_j,\bv_j = (v_{jc},v_{je}),\tau_j).$$

Let $J$ be the total number of subjects ($J=19$ in this case); let $ n^{(t)}_j $ be the number of trials (decisions) made by participant $j$ in condition $t$; denote by $y_{ji}^{(t)}$ the $i^{th}$ decision from subject $j$ under condition $t$. With the usual independence assumptions, the conditional density of all the observations is
$$ p(\by|\bb,\bA,\bv,\bt) = \prod\limits_{j=1}^J \prod\limits_{t \in \{a,n,s\}} \prod\limits_{i=1}^{n^{(t)}_j}  \textrm{LBA}(y_{ji}^{(t)}|b_j^{(t)},A_j,\bv_j,\tau_j), $$
which replaces the generic form in Equation \eqref{eq:p-y-given-alpha-J} with the LBA density of all the observations. Our article makes a small change in the parameterization proposed by \citet{gunawan2020new}. To take into account the constraint that thresholds ($b$) must always be higher than the top of the start point distribution ($A$), we parameterize $c_j^{(t)}=b_j^{(t)}-A_j$ for $j=1,\dots,J;\, t\in \{a,n,s\}$. We follow \citet{gunawan2020new}, and log-transform all the
random effects, which gives them support on the entire real line, and in many cases also leads to approximately normal distributions of the random effects across subjects. For each subject $j=1,\dots,J$, we define the vector of log-transformed random effects
$$\balph_j = (\alpha_{j1},\dots,\alpha_{j7}):= \log \left( c^{(a)}_j,c^{(n)}_j,c^{(s)}_j,A_j,\bv^\top_j = (v_{jc},v_{je}),\tau_j \right).$$
Let $D_{\alpha}$ be the dimension of $\balph_j$ (in this case, $D_{\alpha}=7$). Then, the conditional density of the hierarchical LBA model is defined as $y_{ji}^{(t)}|\balph_j \stackrel{i.i.d.}{\sim}  \textrm{LBA}(y_{ji}^{(t)}|c_j^{(t)},A_j,\bv_j,\tau_j)$ for $j = 1,\dots,J;\, t \in \{a,n,s\}; \, i = 1,\dots,n_j^{(t)}$. The prior for the random effects (that is, the group-level distribution) and the priors for model parameters are as specified in Equations \eqref{eq:prior-random-effects} and \eqref{eq:prior-model-parameters}.	

\subsection{Applying Variational Bayes to the Hierarchical LBA Model}

We first demonstrate the Gaussian VB and Hybrid Gaussian VB methods by using them to estimate the hierarchical LBA model from the data reported by \cite{forstmann2008striatum}. This experiment is small enough to make exact Bayesian inference using MCMC feasible. To assess the quality of the VB approximations, we compare the VB results to the exact posterior estimated using the Particle Metropolis within Gibbs sampler \citep[PMwG:][]{gunawan2020new}.

The posterior was approximated using Gaussian VB and Hybrid Gaussian VB; in each case using 20 factors to reduce the dimension of the approximating distribution. This represents a substantial simplification from the full model, which has $p=161$ parameters (7 group-level mean parameters, 21 parameters for the covariance matrix of those means, and $19\times7$ random effects for individual subjects). The lower bounds and gradients are estimated at each iteration using $N=10$ Monte-Carlo samples. The step sizes are set by using the adaptive learning rate algorithm ADADELTA with $\xi =  10^{-7}$ and $v=0.95$; see Appendix \ref{appendix:VB-details}. The computation time for the Gaussian VB and Hybrid Gaussian VB methods were both less than 5 minutes on an average desktop computer (Intel(R) Core(TM) i5-6500 CPU, 3.20GHz and 8 GB of RAM). By comparison, the run time for the PMwG method on the same system was approximately 2 hours.

\begin{figure}[H]
	\centering
	\hspace*{-2.5cm}\includegraphics[scale = 0.3]{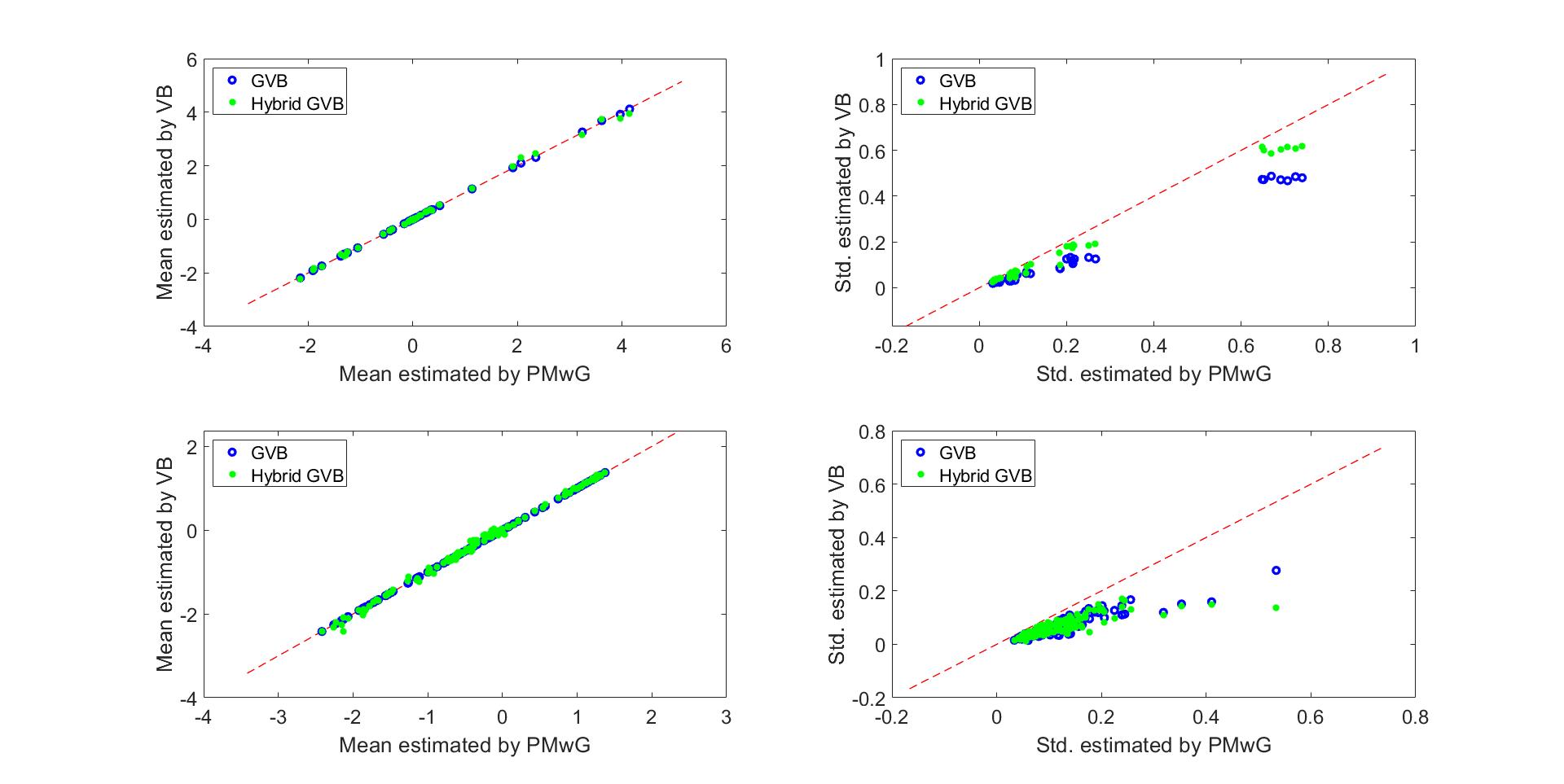}\\
	\caption{Comparing the means and standard deviations of the marginal posterior distributions estimated by VB (vertical axis) against the exact values calculated using PMwG (horizontal axis). The top panels show the means and standard deviations of the group-level parameters. The bottom panels show the means and standard deviations of the random effects. The Gaussian VB (GVB) and Hybrid GVB methods accurately recover the mean of the posterior, but underestimate the standard deviation.}\label{fig:Forstmann_moment_plot}
\end{figure}

Hybrid Gaussian VB provides a better approximation to the posterior distribution, as indicated by a greater lower bound than Gaussian VB (7,275 vs. 7,242). To assess the quality of the marginal inference, the two left panels of Figure \ref{fig:Forstmann_moment_plot} compare the posterior means estimated by the VB methods against the exact posterior means calculated using PMwG. Both Gaussian and Hybrid Gaussian VB capture the posterior means quite precisely, for both the group-level mean parameters (top left panel) and for the individual-subject random effects (lower left panel). The right panels of Figure \ref{fig:Forstmann_moment_plot} shows the corresponding comparison for the estimated standard deviations of the posterior distribution. The standard deviation of the posterior is underestimated by both methods, which is typical for VB. However, Hybrid Gaussian VB provides much more accurate estimates for the posterior standard deviations of the group-level parameters than Gaussian VB (top right panel);  this demonstrates a clear advantage of the Hybrid Gaussian VB method.

\begin{figure}
	\hspace{-1cm}\includegraphics[scale=0.25]{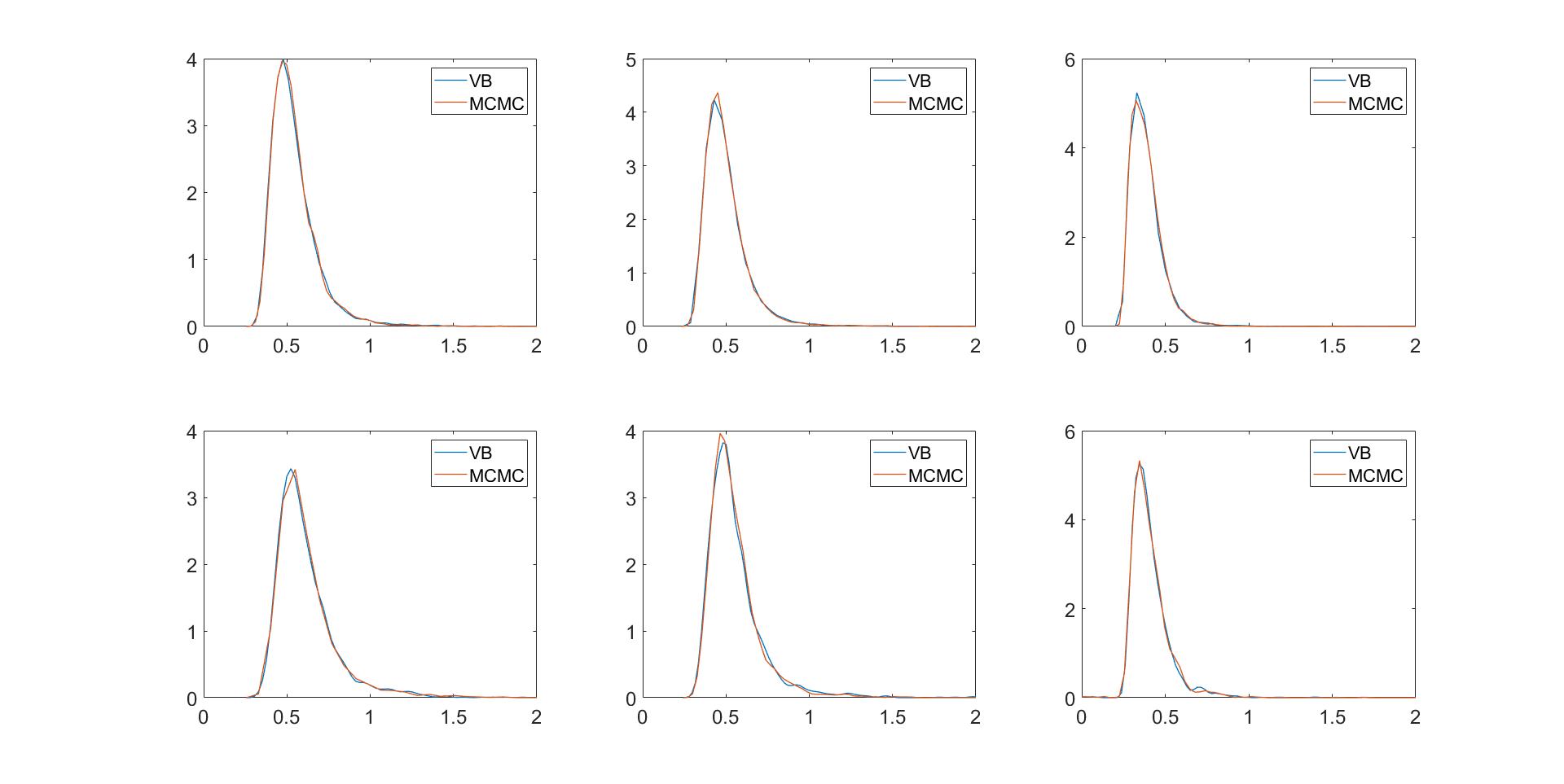}
	\caption{Posterior predictive response time distributions for participant 2. The top panels show correct responses and the bottom panels show incorrect responses. Columns show the emphasis conditions: accuracy (left), neutral (middle) and speed (right).}
	\label{fig:post-pred-subject 2}
\end{figure}

We now compare the predictive densities estimated using PMwG with ones obtained with the hybrid VB approximation.
Figure \ref{fig:post-pred-subject 2} shows these posterior predictive densities for subject 2; results for other subjects are similar. The fact that the posterior predictive densities are very well approximated by VB supports the claim that VB gives very good predictions. Appendix \ref{appendix:posterior-predictive-densities} gives the algorithm to obtain the predictive densities for the hierarchical LBA models.

\section{CVVB in Action: A More Thorough Evaluation of Selective Influence in LBA Models of Decision-Making}

The notion of ``selective influence'' has been important in evaluating psychological models, including evidence accumulation models \citep{RatcliffRouder1998,VossEtAl2004}. An experimental manipulation (e.g., changing the brightness of a perceptual decision stimulus) is said to selectively influence a particular model parameter (e.g., drift rate) if the model can account for differences in observed data caused by the manipulation via adjustments in only that one parameter.

\cite{rae2014hare} and \cite{StarnsRW2012} identified an important violation of selective influence in both the LBA model and the related diffusion decision model. When decision-makers were asked to adjust their speed-accuracy tradeoff strategies, the models required more than just changes in threshold parameters to explain the observed data. Instead, the models required changes in threshold parameters and drift rate parameters -- contrary to expectation, the speed-accuracy tradeoff manipulation did not selectively influence threshold parameters.

\cite{rae2014hare} and \cite{StarnsRW2012} carried out  inference about the model parameters using statistical methods which were available to them at the time. The methods presented here allow these results to be improved in important ways. Firstly, the models can be treated using a random effects structure, which allows for person-to-person variation. Secondly, using the CVVB method, a much more complete set of candidate model parameterizations can be investigated. This reduces the dangers posed by experimenter bias. Below, we update those earlier findings by reanalysing three previously-reported data sets, using three very different decision-making tasks. In each case, we investigate the question of selective influence by enumerating a comprehensive set of models for comparison, using CVVB to choose between them. Before reanalysing the real data, we present a simulation study which shows the properties of our methods.

\subsection{Case Study 1: The Speed-Accuracy Tradeoff in Perceptual Decisions}

As the first demonstration, we reconsider the experiment conducted by \citet{forstmann2008striatum}. In our earlier application of VB methods to this data set, we made the standard selective influence assumption: the effect of the speed-accuracy tradeoff manipulation is entirely explained by separate response threshold settings ($c$) for the speed, neutral and accuracy emphasis conditions, with all remaining random effects, i.e., subject-level parameters, estimated to common values across conditions. Whether selective influence of this manipulation holds in the LBA model parameters can be tested by investigating whether different threshold settings are required for the different conditions, and/or whether other random effects are also required to be different across those conditions, particularly the drift rates, $\bv$. We investigated a set of 27 different models, ranging in complexity from a null model (the random effects are the same across conditions) through to a very complex model with three random effects for $\tau$, three for threshold $c$, and three pairs of drift rates $\bv$. Each model is denoted by the number of random effects for $c,\bv$ and $\tau$. For instance, model 3-2-1 denotes an LBA model with 3 random effects for thresholds ($c^{(n)},c^{(s)},c^{(a)}$), 2 random effects for drift rates ($\bv_1,\bv_2$), and only 1 random effect for non-decision time ($\tau$).

\subsubsection{Simulation Study}

We first conducted a simulation study to investigate the performance of the CVVB procedure, and in particular its ability to detect the data generating model. The simulation design is based on \citet{forstmann2008striatum} experiment with 19 participants and 1,000 trials per participant, where the data generating process is an LBA model. The data generating (``true'') model parameters $\bmualph$ and $\bSigalph$ are set to estimated from the data using PMwG for model 3-1-1, which is the selective influence model, i.e., three threshold settings for the three conditions, but no change in the other parameters. We ran 100 independent replications, and in each replication, we repeated the following steps for each of the $j=1,\dots,19$ simulated participants:
\begin{enumerate}
	\item Sample $\balph_j\sim N(\bmualph,\bSigalph)$
	\item Transform $\balph_j$ back to the natural parameterization $(b^{(a)}_j,b^{(n)}_j,b^{(s)}_j,A_j,v_j,\tau_j)$.
	\item Simulate 1,000 trials for subject $j$ as follows
	\begin{itemize}
		\item Sample 350 pairs $(RT_{ij},RE_{ij})\sim \textrm{LBA}(t,c|b^{(a)}_j,A_j,v_j,\tau_j)$.
		\item Sample 350 pairs $(RT_{ij},RE_{ij})\sim \textrm{LBA}(t,c|b^{(n)}_j,A_j,v_j,\tau_j)$.
		\item Sample 300 pairs $(RT_{ij},RE_{ij})\sim \textrm{LBA}(t,c|b^{(s)}_j,A_j,v_j,\tau_j)$.
	\end{itemize}
\end{enumerate}

\begin{figure}
	\centering
	\includegraphics[scale = 0.3]{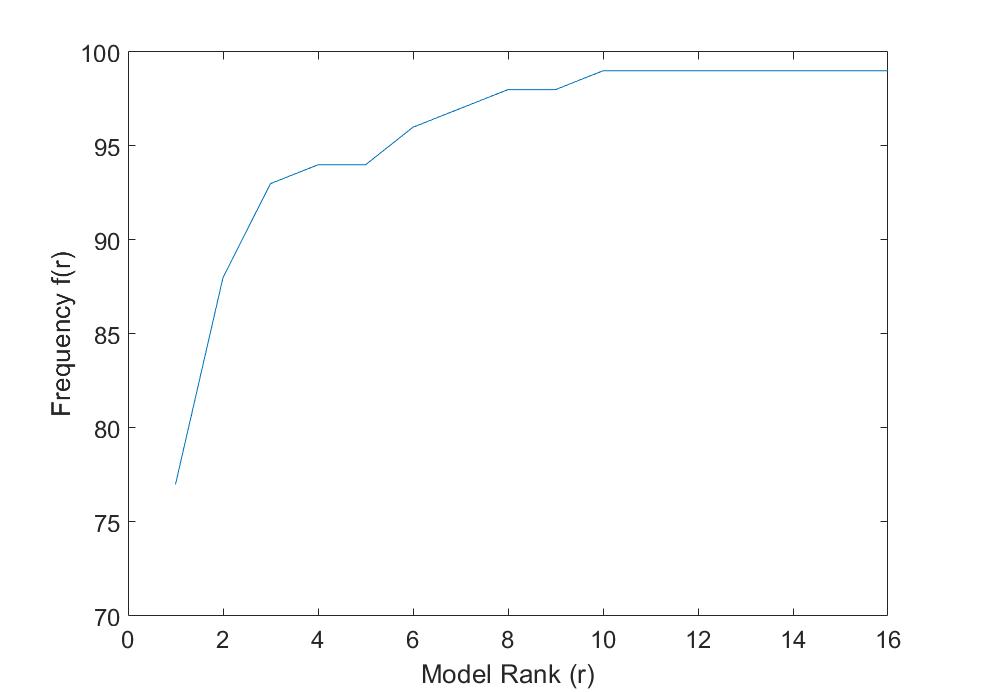}
	\caption{Sensitivity of the CVVB procedure for data simulated from \citet{forstmann2008striatum} design. The $y$-axis shows the frequency (from 100 replications) with which the data-generating model is ranked in the top $r$ models ($x$-axis). The best model is ranked 1, the second best model is ranked 2, and so on through to the worst model which is ranked 27. A procedure with high sensitivity has large $f(r)$ ($y$-axis) for small model ranks ($x$-axis).}	\label{fig:sensitivity_functions}
\end{figure}

For each of the 100 simulated data sets, we used $5$-fold CVVB to estimate all 27 candidate LBA models and then ranked the models using ELPD. Figure \ref{fig:sensitivity_functions} shows the sensitivity of the CVVB procedure: the number of times out of 100 replications that the data-generating model was ranked in the top $r$ models ($x$-axis). For example, the data-generating model was ranked amongst the top 3 candidates in 94 of the 100 replications, and was correctly ranked as the most likely model over 75\% of the time. Given the small size of the simulated data sample ($n=19$ subjects) and the approximate nature of the CVVB algorithm, we consider this as good performance. Of particular importance is that the data-generating model was quite simple relative to some of the candidate models, indicating that the CVVB procedure appropriately manages the flexibility of the set of models under consideration.

\subsubsection{Analysis of the Real Data}

The performance of all 27 candidate models in the \citet{forstmann2008striatum} data was evaluated using CVVB, Hybrid Gaussian VB, 5 folds, and 15 factors to reduce the dimension of the approximating distribution. We compared ELPD estimated by CVVB with the marginal likelihood estimated by the Importance Sampling Squared (IS$^2$) method of \cite{tran2020robustly}.
Table~\ref{table:Forstmann_log_score} compares the estimated marginal likelihood for each model (right-most column) against the $\elpdcvvb$ (second-to-right column). The left-most column gives each model an index number, which we use in the plots below. There is general agreement between the CVVB method and the corresponding marginal likelihood estimate from the exact method. For example, both methods place the same three models (11, 22, and 23) among their top four best models. The 12 worst-ranked models by the two methods are also the same.

\begin{figure}
	\centering
	\hspace*{-1cm}\includegraphics[scale = 0.25]{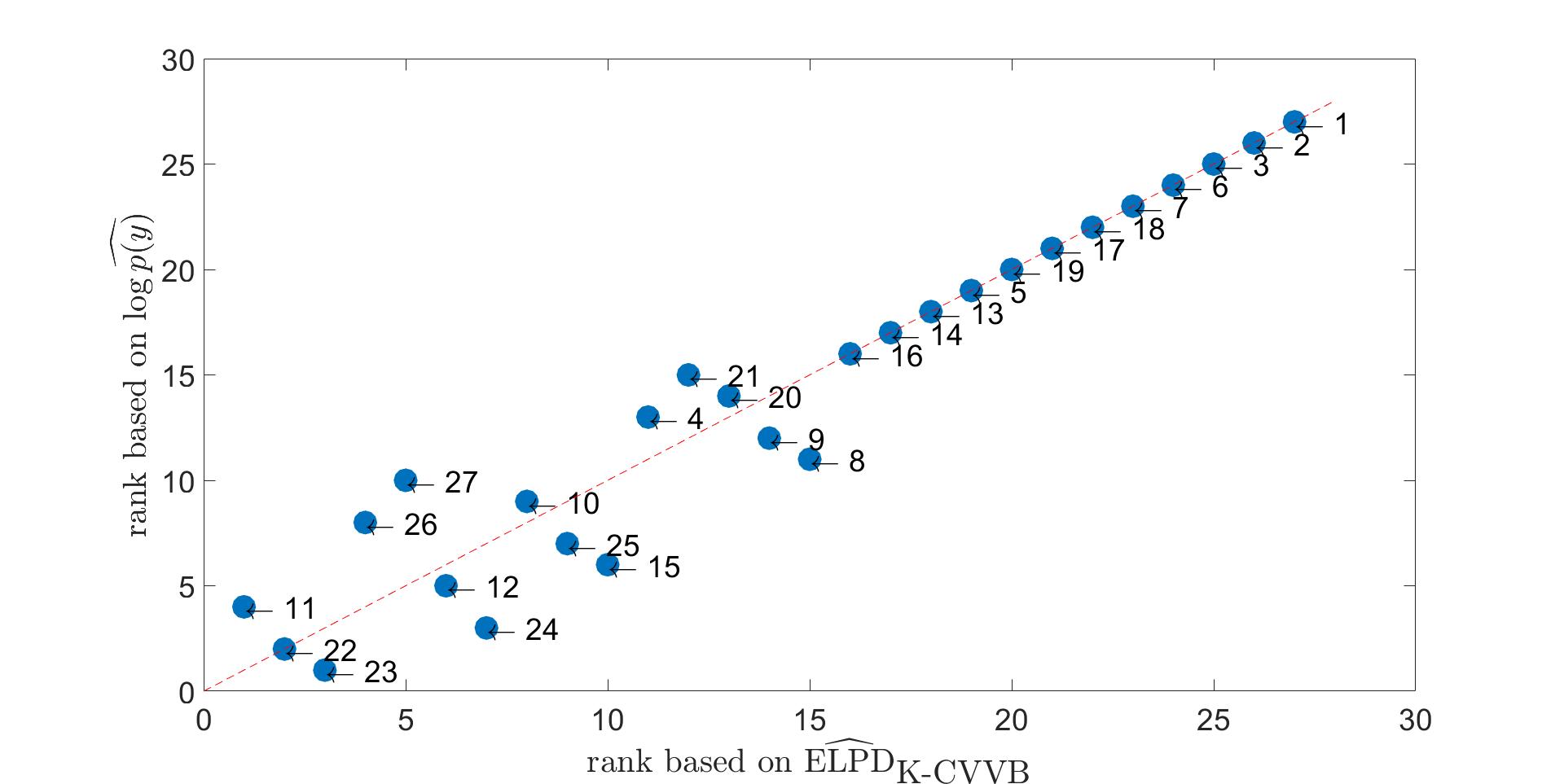}
	\caption{CVVB model ranks ($x$-axis) plotted against marginal likelihood model ranks ($y$-axis) for \citet{forstmann2008striatum} data.}\label{fig:Forstmann_rank_plot}
\end{figure}

Figure~\ref{fig:Forstmann_rank_plot} compares the ranking on the set of 27 models implied by CVVB with the ranking implied by marginal likelihood. While there are some differences evident in the rankings given to middle-ranked models, overall the agreement is quite good. The Spearman rank correlation of the rankings implied by the two model selection methods is $\rho = .9602$. Both model selection approaches agree on the central conclusion: that the speed-accuracy manipulation did not selectively influence threshold parameters. The top-ranked models in both analyses include effects of the speed-accuracy manipulation on drift rates and/or non-decision times, in addition to threshold settings.

\begin{table}
	\caption{Model selection via CVVB and marginal likelihood for the 27 LBA models fitted to data reported by \citet{forstmann2008striatum}. The last column lists the log-marginal likelihood estimated by the IS$^2$ method with standard errors in brackets.\label{table:Forstmann_log_score}}
	\centering%
	\begin{tabular}{c|c|c|c}
	\hline\hline
		Model& Model &  $\elpdcvvb$ & $\logmarg$\\
		Index&($c-\bv-\tau$) &  & (IS$^2$ method)\\
		\hline	
		11 &	2-3-2 	& 1,548.9 	& 7,584.9 (0.2)\\
		22 &	3-2-3	& 1,548.6 	& 7,591.2 (0.2)\\	
		23 &	3-2-2	& 1,548.0 	& 7,595.3 (0.2)\\
		26 &	3-3-2	& 1,547.4 	& 7,572.1 (0.3)\\
		27 &	3-3-3	& 1,547.2 	& 7,556.7 (0.4)\\
		12 &	2-3-1 	& 1,545.7 	& 7,580.1 (0.6)\\
		24 &	3-2-1	& 1,540.5 	& 7,591.2 (0.2)\\
		10&	2-3-3	 &	 1,536.5 &	 7,571.4 (0.3)\\
		25 &	3-3-1	& 1,536.2 	& 7,573.4 (1.4)\\
		15 &	2-2-3	& 1,535.6 	& 7,573.6 (0.5)\\
		4 &	1-2-3	 &	 1,530.6 &	 7,536.4 (0.2)\\
		21 &	3-1-3	& 1,530.4 	& 7,527.8 (0.1)\\
		20 &	3-1-2	& 1,528.6 	& 7,529.9 (0.1)\\
		9 &	1-3-3	 &	 1,527.4 &	 7,537.3 (0.5)\\
		8 &	1-3-2	 &	 1,527.1 &	 7,541.6 (0.5)\\
		16 &	2-1-3	& 1,525.2 	& 7,510.8 (0.2)\\
		14 & 	2-2-2	& 1,523.1 	& 7,508.4 (0.3)\\
		13 & 	2-2-1	& 1,522.2 	& 7,500.5 (0.1)\\
		5 &	1-2-2	 &	 1,516.7 &	 7,465.9 (0.2)\\
		19 &	3-1-1	& 1,516.2 	& 7,461.8 (0.7)\\
		17 &	2-1-2	& 1,512.6 	& 7,444.0 (0.1)\\
		18 &	2-1-1	& 1,493.2 	& 7,359.0 (0.1)\\
		7 &	1-3-1	 &	 1,473.9 &	 7,279.1 (0.3)\\
		6 &	1-2-1	 &	 1,462.7 &	 7,220.8 (0.1)\\
		3 &	1-1-3	 &	 1,433.2 &	 7,028.9 (0.1)\\
		2 &	1-1-2	 &	 1,413.7 &	 6,947.0 (0.1)\\
		1 &	1-1-1	 &	 1,060.4 &	 5,199.5 (0.1)\\	\hline\hline
	\end{tabular}
\end{table}

\subsection{Case Study 2: The Speed-Accuracy Tradeoff in Recognition Memory}

\cite{rae2014hare} reported a new experiment to test selective influence in a decision-making task based on memory recognition (as opposed to perceptual discrimination, as above). For this, 47 participants were asked to study lists of words and then repeatedly decide whether given prompt words were old (from the studied lists) or new (not). For some decisions, participants were instructed to respond very urgently (speed emphasis) and for others to respond very carefully (accuracy emphasis).

To evaluate the selective influence of the speed/accuracy manipulation on the threshold parameters, we investigated a large set of LBA models. We allowed the random effects for the threshold ($c$) to vary between response accumulators (``old'' vs. ``new'') in order to capture the biases in different subject's responding patterns. We also allowed drift rates ($\bv$) to vary between accumulators and according to whether the stimulus was actually an old or new word, which captures the basic ability of subjects to do the memory task. This investigation compares the 16 models given in Table \ref{table:case 2 models}. In the table, models are numbered from 1 (the simplest) to 16 (the most complex). For this data set, and the following one, we have adopted a notation based on the experimental manipulations to describe the models. For example, the notation $E*R$ in the second column indicates that the corresponding parameter for that column ($c$) is allowed to vary with both the response accumulator (R) and the speed vs. accuracy emphasis manipulation (E). The letter ``S'' indicates the manipulation of studied (old) vs. not studied stimulus words, and the letter ``M'' indicates the match between the stimulus class and the response. A parameter is indicated by $1$ if it is common across conditions. For example, in model 1, we allow $c$ to vary with the response accumulator R, $v$ to vary with the stimulus S and the stimulus-accuracy match M; $s$ is only affected by the stimulus M, and both $A$ and $\tau$ are common across accumulators and conditions.

\begin{table}
	\caption{Model selection via CVVB and marginal likelihood for the 16 LBA models fitted to the data reported by \citet{rae2014hare}. The last column lists the log-marginal likelihood estimated by the IS$^2$ method with the standard errors in brackets.}\label{table:case 2 models}
	\centering
	\hspace*{-0.5cm}	\begin{tabular}{c|c|c|c|c|c|c|c}
		Model	&	\multicolumn{5}{c|}{Model}   & $\elpdcvvb$ & $\logmarg$\\
		Index & $c$ & $A$ & $v$ & $s$ & $\tau$ &  & (IS$^2$ method)\\		
		\hline
		16 &	$E*R$ & $E$ & $E*S*M$ & $M$	& $E$ & 1,190.5&  5,944.4 (0.6)\\
		12 &	$E*R$ & $1$ & $E*S*M$ & $M$	& $E$ & 1,174.5 & 5,907.1 (1.2)\\
		15 &	$E*R$ & $E$ & $E*S*M$ & $M$	& $1$ & 1,172.2 & 5,942.7 (0.9)\\
		8 &	$R$ & $E$ & $E*S*M$ & $M$	& $E$ & 1,165.1 & 5,861.7 (0.2)\\
		11 &	$E*R$ & $1$ & $E*S*M$ & $M$	& $1$ & 1,159.2 & 5,894.8 (0.7)\\
		13 &	$E*R$ & $E$ & $S*M$ & $M$	& $1$ & 1,142.6 & 5,570.7 (1.2)\\
		4 &	$R$ & $1$ & $E*S*M$ & $M$	& $E$ & 1,127.1 & 5,830.8 (1.9)\\
		14 &	$E*R$ & $E$ & $S*M$ & $M$	& $E$ & 1,112.3 & 5,574.0 (0.7)\\
		10 &	$E*R$ & $1$ & $S*M$ & $M$	& $E$ &  1,105.3 & 5,490.6 (2.8)\\
		9 &	$E*R$ & $1$ & $S*M$ & $M$	& $1$ & 1,094.5 & 5,404.6 (0.2)\\
		3 &	$R$ & $1$ & $E*S*M$ & $M$	& $1$ & 1,053.1 & 5,499.5 (0.5)\\
		6 &	$R$ & $E$ & $S*M$ & $M$	& $E$ & 1,052.1 & 5,238.4 (0.5)\\
		7 &	$R$ & $E$ & $E*S*M$ & $M$	& $1$ & 1,041 & 5,581.9 (0.3)\\
		5 &	$R$ & $E$ & $S*M$ & $M$	& $1$ & 851.6 & 4,308.9 (0.5)\\		
		2 &	$R$ & $1$ & $S*M$ & $M$	& $E$ & 758.5 & 3,793.6 (0.2)\\
		1 &	$R$ & $1$ & $S*M$ & $M$	& $1$ & -574.5 & -3,026.8 (0.3)\\		
		\hline\hline
	\end{tabular}
\end{table}

\begin{figure}
	\centering
	\hspace*{-1cm}\includegraphics[scale = 0.25]{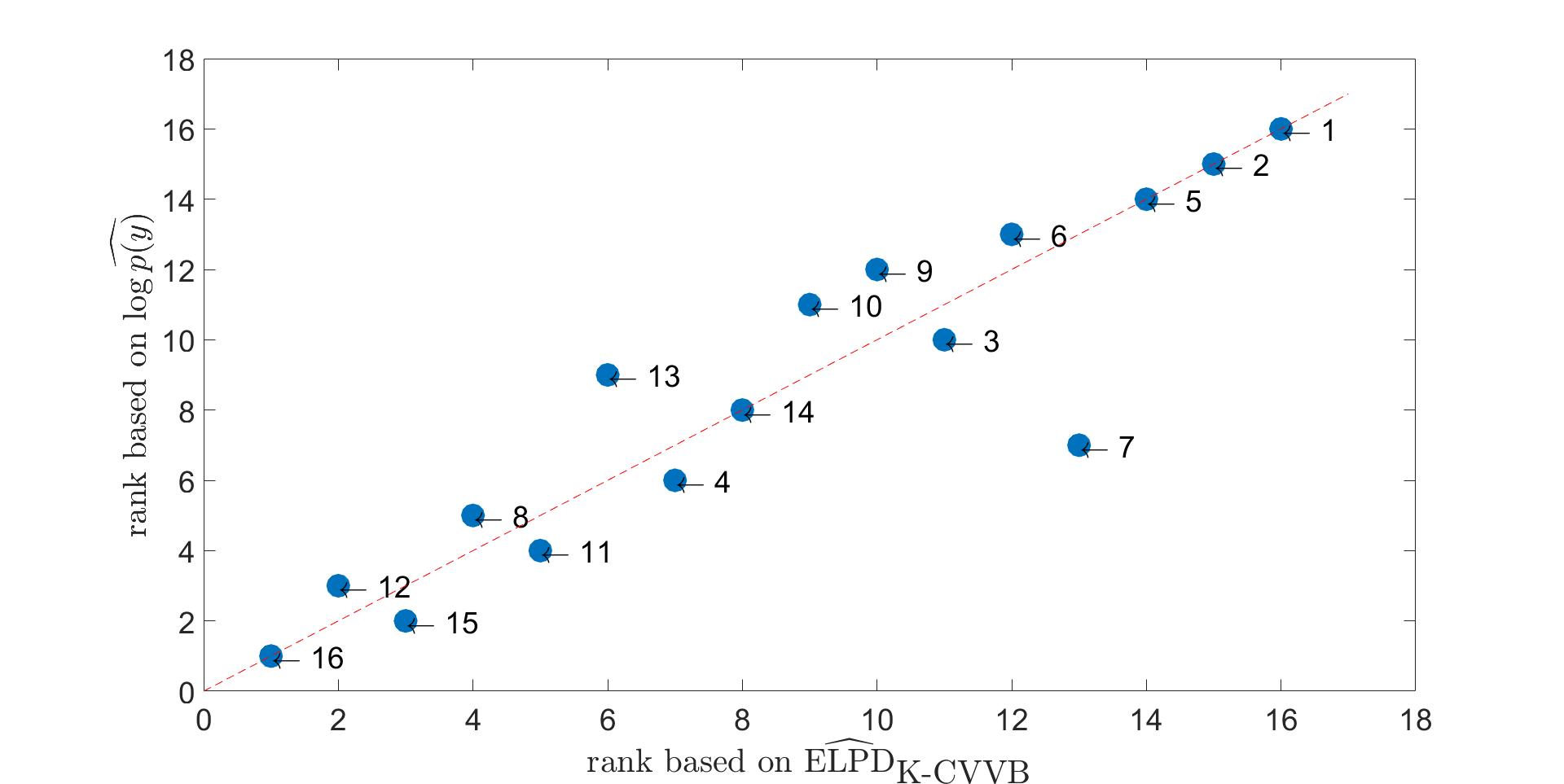}
	\caption{CVVB model ranks ($x$-axis) plotted against marginal likelihood model ranks ($y$-axis) for \citet{rae2014hare} data.}\label{fig: LB_plot_1}
\end{figure}

Table~\ref{table:case 2 models} compares ELPD (estimated using CVVB) with marginal likelihood (estimated using IS$^2$). The two model selection methods are quite consistent in this example, agreeing on the same set of five best-ranked models and four out of the five worst-ranked models. Figure~\ref{fig: LB_plot_1} compares the rankings implied by the two methods, and, once again, the agreement is quite good (Spearman rank correlation of $\rho=.9118$). As for Case Study 1, both methods agree on the primary conclusion: that the speed/accuracy manipulation did not selectively influence threshold parameters. For both model selection methods, the top 5 ranked models all include effects of the speed/accuracy manipulation (``E'' in Table~\ref{table:case 2 models}) on parameters other than thresholds ($c$).

\subsection{Case Study 3: The Speed-Accuracy Tradeoff in Lexical Decisions}

The first two case studies investigated the selective influence of speed/accuracy manipulations on the threshold parameter of the LBA model in perceptual decisions \citep{forstmann2008striatum} and mnemonic decisions \citep{rae2014hare}. The third case study extends the analysis to a different decision-making domain: lexical decisions. In addition, this third case study emphasizes the benefit of VB methods because the set of models to be compared is much larger (256). Model comparison using exact methods such as MCMC with such a large class of models is very expensive.

The lexical decision task is a commonly used method for studying highly-practiced processes in reading. Participants are required to rapidly decide whether strings of letters are either valid English words (e.g., ``WORD'') or non-words (e.g., ``WERD''). We analyze data from Experiment~1 of \citet{wagenmakers2008diffusion}. In this experiment, 17 native English speakers made lexical decisions and were sometimes instructed to respond as quickly as possible (speed emphasis) and sometimes to respond as accurately as possible (accuracy emphasis). In addition, there were three different kinds of words used, which changed the difficulty of the decision. Some words were very common words (high frequency), such as ``CARS''. Others were uncommon words (low frequency), such as ``COMB'', and others were very-low frequency words, such as ``DALE''. Participants find it more difficult to distinguish between very low frequency words and non-words.

We use $E$ to represent the speed/accuracy conditions, $C$ for the responses (error (e) or correct (c)), and $W$ for the four levels of word frequency (high frequency, low frequency, very low frequency or non-word). The performance of 256 models was evaluated. The simplest model allows only the mean drift rate to differ between correct and error accumulators ($c\sim 1,\ A\sim 1,\ v\sim C,\ s \sim 1,\ \tau \sim 1$), reflecting the idea that participants could perform the basic lexical decision task (i.e., separate words from non-words) but the other manipulations had no effects. The most complex model allows for effects on many different parameters ($c\sim C*E,\ A\sim C*E,\ v\sim E*W*C,\ s \sim 1,\ \tau \sim E$).

With a large number of competing models, model selection based on the log marginal likelihood is extremely costly -- this is one of the primary reasons for using VB methods. Therefore, we did not estimate the marginal likelihood for all the models. Instead, we propose a mixed approach in which we use CVVB to quickly screen through all the models. This results in an approximate ranking for all the models in approximately 16 hours. From this ranking, we selected a small subset (the best 10 and the worst 10) for follow-up using slower exact methods to estimate the posterior distributions and marginal likelihood.

\begin{figure}
	\centering
	\hspace*{-1cm}	\includegraphics[scale = 0.25]{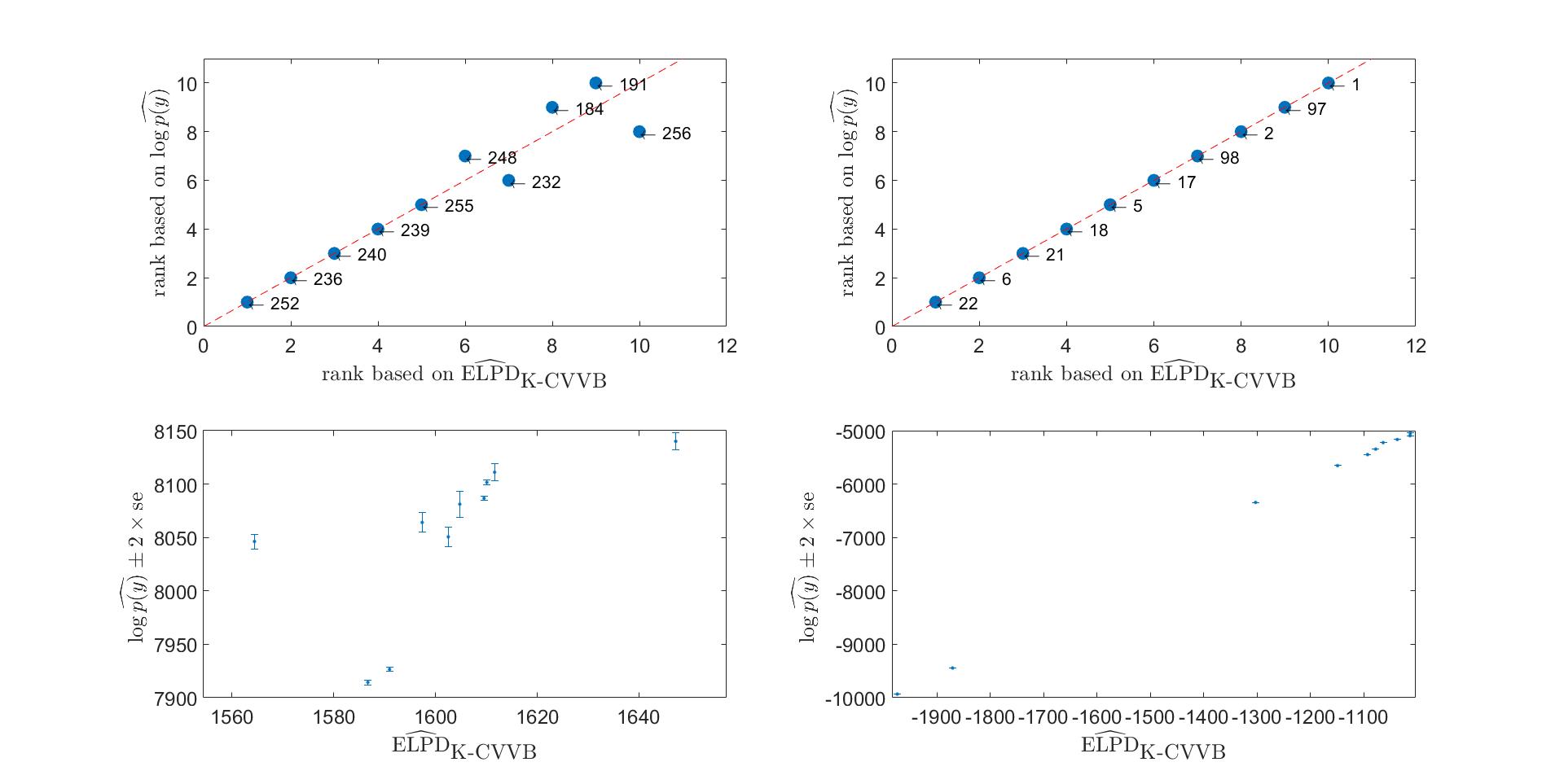}
	\caption{$\elpdcvvb$ and marginal likelihood estimates for the 10 best models (left panels) and the 10 worst models (right panels) for the data reported by \citet{wagenmakers2008diffusion}. The lower panels plot the $\elpdcvvb$ ($x$-axes) against the marginal likelihood estimate ($\pm$ two standard errors; $y$-axes). The upper panels show the corresponding model ranks from the two methods.}\label{fig: Lexical_scatterplot_rankings}
\end{figure}

Table~\ref{table: Lexical} lists the results for just these selected best models, comparing ELPD (estimated using CVVB) with marginal likelihood (estimated using IS$^2$). Figure~\ref{fig: Lexical_scatterplot_rankings} compares the $\elpdcvvb$ with the log marginal likelihood both in absolute terms (lower panels) and rankings (upper panels). The figure shows the comparison for both the 10 best models according to $\elpdcvvb$ (left panels) and the 10 worst (right panels). For the 10 best models, the two methods closely agree on both the relative ranking of the models (Spearman rank correlation of $\rho =0.9515$) and even the distances between them in terms of predictive performance, with the possible exception of the most complex model (256). The agreement is even better for the 10 worst models.

As for Case Studies 1 and 2, the new analysis confirms the earlier results that the speed/accuracy manipulation does not selectively influence the threshold parameters. All of the 10 best models (top half of Table~\ref{table: Lexical}) include effects of the speed/accuracy manipulation (``E'') on parameters other than the threshold (column $c$).

\begin{table}
	\caption{Model selection via CVVB and marginal likelihood for the 10 best models (above the solid line) and the 10 worst models (below the solid line) fitted to the data reported by \citet{wagenmakers2008diffusion}. The last column lists the log-marginal likelihood estimated by the IS$^2$ method with the standard errors in brackets.}\label{table: Lexical}
	\centering%
	\begin{tabular}{c|c|c|c|c|c|c}
		Model	&	\multicolumn{4}{c|}{Model}  &  \multicolumn{1}{c|}{$\elpdcvvb$} & $\logmarg$\\
		Index & $c$ & $A$ & $v$ &  $\tau$ &  & (IS$^2$ method) \\	
		\hline
		252 & C & C*E & C*W*E & E & 1,647.2  &  8,140.0 (3.9)\\
		
		236 & C & C & C*W*E &  E & 1,611.6 & 8,111.1 (4.0) \\
		
		240 & C*E & C & C*W*E & E & 1,610.1 &  8,101.6 (1.3)\\
		
		239 & C*E & C & C*W*E &  1 & 1,609.5 & 8,086.7 (1.1) \\
		
		255 & C*E & C*E & C*W*E & 1 & 1,604.8 & 8,081.2 (6.1)\\
		
		248 & C*E & E & C*W*E & E & 1,602.5  & 8,050.6 (4.6) \\
		
		232 & C*E & 1 & C*W*E &  E & 1,597.4 & 8,064.1 (4.5) \\
		
		184 & C*E & E & C*W & E & 1,590.9  & 7,926.5 (1.0) \\
						
		191 & C*E & C*E & C*W & 1 & 1,586.7 &  7,914.4 (1.1)\\
		
		256 & C*E & C*E & C*W*E & E & 1,564.4  & 8,046.2 (3.4)\\		

		\hline
		\hline
		
		22 & E & E & 1 & E & -1,015.3  &  -5,043.2 (0.6)\\
		
		6 & E & 1 & 1 & E & -1,007.0  &  -5,094.4 (0.2)\\
		
		21 & E & E & 1 & 1 & -1,017.8  &  -5,165.9 (0.6)\\
		
		18 & 1 & E & 1 & E & -1,028.0  &  -5,223.9 (0.1)\\
		
		5 & E & 1 & 1 & 1 &  -1,065.0  &  -5,346.0 (0.1)\\
		
		17 & 1 & E & 1 & 1 &  -1,085.0  &  -5,448.2 (0.1)\\
		
		98 & 1 & 1 & W & E & -1,144.2  &  -5,654.4 (0.2)\\
		
		2 & 1 & 1 & 1 & E & -1,274.3  &  -6,343.5 (0.05)\\
		
		97 & 1 & 1 & W & 1 & -1,867.4  &  -9,444.7 (0.3)\\
		
		1 & 1 & 1 & 1 &  1 & -1,969.9 & -9,932.4 (0.2) \\
		
	\end{tabular}
\end{table}

\section{Discussion}

This paper proposes Hybrid VB method for approximate Bayesian inference with psychological models; it
is more efficient than previous VB methods for such models. The performance of the VB method is demonstrated with applications in decision making. An important development from our work is the coupling of VB methods for model estimation with cross-validation methods for model selection. The combined CVVB approach is a computationally efficient method for model selection. This method is particularly useful when the number of models to be compared is large, which can make exact methods (such as MCMC) infeasible. Our simulation study shows that CVVB accurately identifies the data-generating model, and our analyses of real data repeatedly demonstrate that the CVVB results agree closely to model selection by marginal likelihood, estimated by exact (i.e., simulation consistent) algorithms. However, some users may still want to base their final conclusions on exact methods, and for that situation we propose using CVVB as a model screening tool. CVVB can be used to efficiently ``screen'' a large set of models, and quickly identify a much smaller number of candidates for follow-up by slower, exact methods. The CVVB method allows a more thorough investigation of an important question about ``selective influence'' in the speed-accuracy tradeoff than previous approaches.


VB methods have already been used in other domains of psychological research as a fast alternative to MCMC, but mostly in much simpler models than here. For instance, VB methods have been used to study the impact of three prior distributions on Bayesian parameter recovery in very simple models, with just one or two parameters. In most of these simple cases the authors found VB to be both fast and also highly accurate, and recommend VB for use with hierarchical models in particular because the method is computationally effective, quick, and accurate. Beyond parameter recovery exercises, VB has also been used to investigate probabilistic cognitive models of how people represent temporal structure in the world \citep{markovic2019predicting}, and to approximate solutions to the inverse Bayesian decision theory problem in the context of learning and decision-making \citep{daunizeau2010observing}.

While these applications of VB are interesting and effective, they all employ the so-called ``mean field VB'', which assumes a simplified factorization for the variational distribution $q$. Mean field VB ignores the posterior dependence between the blocks of model parameters, and requires analytical calculation of model-specific expectations \citep{ormerod2010explaining}. These can be challenging to compute, or simply unavailable, for many interesting psychological models . This has been a major hurdle to the uptake of VB for substantively interesting psychological models.

By contrast, the ``fixed form'' VB method we have used is more flexible and widely applicable. It takes into account the posterior dependence between the model parameters and does not require any calculation of model-specific expectations. In recent work promoting the use of VB, \cite{galdo2019variational} also proposed fixed form VB -- their methods adopt an optimization strategy called differential evolution to bypass the need to compute analytical expectations. Nevertheless, Galdo et al. still assume a simplified factorization structure for the variational distribution $q$, and thus do not account for posterior dependence between blocks of parameters. \citet{galdo2019variational} test their approach using two benchmark cognitive models, a non-hierarchical (single subject) LBA model and a hierarchical version of Signal Detection Theory. Our work extends that of Galdo et al. in at least two important aspects. First, it examines for hierarchical LBA models with a more complete parameterization. The multivariate Gaussian group-level distribution accounts for between-subject differences and also for the correlation of the random effects, and therefore provides a more realistic representation of prior knowledge. Second, our fixed-form VB approach takes into account the dependence structure of the posterior and incorporates some of the latest advances in the fixed form VB literature.

We hope that the VB methods developed in the article will be taken up and extended by other researchers. To assist in this, we have written a user manual document with clear instructions to show the interested readers how to modify our codes for their analyses. We share the code, data and the user manual online, at \href{https://github.com/Henry-Dao/CVVB}{https://github.com/Henry-Dao/CVVB}. A document with clear instructions is provided. The methods developed in this paper are quite general and are not limited to the LBA model. Our approach will translate easily to other cognitive psychology models provided the group-level models are maintained with relatively uninformative priors. 

\section{Acknowledgements} 
The research of Viet Hung Dao, Minh Ngoc Tran, Robert Kohn and Scott Brown was partially supported by ARC Discovery grant DP180102195. 

\bibliography{myreferences}
\newpage
\appendix

\section{\bf Variational Bayes Details}\label{appendix:VB-details}
\subsection{Details of the Optimization Methods}	

We use gradient-based search methods to maximize the lower bound, which require computing $\gradlb$, the gradient of $\lb$ with respect to the variational parameters $\blamb$. In most cases it is impossible to compute $\gradlb$ analytically, but $\gradlb$  can be estimated unbiasedly. For this reason, stochastic gradient ascent methods \citep{robbins1951stochastic} are often used to optimize $\lb$. These methods start from some initial value $\blamb^{(0)}$ for $\blamb$ and update it recursively by following the gradient vector ``uphill'':
\begin{equation}
	\label{eqn:SGD}
	\blamb^{(t+1)} = \blamb^{(t)} + \boldsymbol{\rho}_t\odot\widehat{\nabla_{\blamb}\mathcal{L}(\blamb^{(t)})},
\end{equation}
where $\brho_t$ is a vector of step sizes, $\odot$ denotes the element-wise product of two vectors, and $\gradlbest$ is an unbiased estimate of the gradient of $\gradlb$.

\subsubsection{A ``reparameterization trick''}
The performance of stochastic gradient ascent depends greatly on the variance of the noisy gradient estimate $\gradlbest$. Performance can therefore be improved by employing variance reduction methods. A popular variance reduction method  is the so-called ``reparameterization trick'' \citep{kingma2013auto,rezende2014stochastic}. If we can write $\btheta\sim \qvb$ as $\btheta = u(\beps;\blamb)$ with $\beps \sim f_{\epsilon}$ which does not depend on $\blamb$, then the lower bound and its gradient can be written as the expectations
\begin{align}\label{eq: gradlb}
	\lb &= E_{f_{\epsilon}}\left[ \log p(\by,u(\beps;\blamb)) - \log q_{\blamb}(u(\beps;\blamb)) \right],\notag\\
	\gradlb &= E_{f_{\epsilon}}\bigg[ \nabla_{\blamb} u(\beps;\blamb) \left[\nabla_\theta \log \joint -  \nabla_\theta \log \qvb \right] \bigg].
\end{align}
By sampling $\beps\sim f_{\epsilon}$, it is straightforward to obtain the unbiased estimates of the lower bound and its gradient
\begin{align}\label{eq: gradlb_hat}
	\lbest& := \dfrac{1}{N}\sum\limits_{i=1}^N\left[ \log p(\by,u(\beps^{(i)};\blamb)) - \log q_{\blamb}(u(\beps^{(i)};\blamb)) \right],\notag\\
	\gradlbest& := \dfrac{1}{N}\sum\limits_{i=1}^N\bigg[ \nabla_{\blamb} u(\beps^{(i)};\blamb) \left(\nabla_\theta \log p(\by,\btheta^{(i)}) -  \nabla_\theta \log q_{\blamb}(\btheta^{(i)}) \right) \bigg],
\end{align}
with $\beps^{(i)} \sim f_{\epsilon}, i = 1,\dots,N$. We used $N=10$ in our applications.

\subsubsection{Learning rates and stopping rule}
The elements of the vector $\blamb$ may need very different step sizes (learning rates) during the search, to account for scale or the geometry of the space. We set the step sizes adaptively using the ADADELTA method \citep{zeiler2012adadelta}, with different step sizes for each element of $\blamb$. At iteration $t+1$, the $i$th element $\lambda_i$ of $\blamb$ is updated as
$$ \lambda_i^{(t+1)}=\lambda_i^{(t)}+\Delta\lambda_i^{(t)}.$$
The step size $\Delta\lambda_i^{(t)}:=\rho_i^{(t)}g_{\lambda_i}^{(t)}$, where $g_{\lambda_i}^{(t)}$ denotes the $i$th component of $\widehat{\nabla_{\blamb}\mathcal{L}(\boldsymbol{\lambda}^{(t)})}$ and
$$ \rho_i^{(t)} := \dfrac{\sqrt{E(\Delta_{\lambda_i}^2)^{(t-1)} + \xi}}{\sqrt{E(g_{\lambda_i}^2)^{(t)} + \xi}},$$
where $\xi$ is a small positive constant, with
\begin{align*}
	E(\Delta_{\lambda_i}^2)^{(t)} &= v E(\Delta_{\lambda_i}^2)^{(t-1)} + (1-v)(\Delta_{\lambda_i}^{(t)})^2,\\
	E(g_{\lambda_i}^2)^{(t)} &= v E(g_{\lambda_i}^2)^{(t-1)} + (1-v)(g_{\lambda_i}^{(t)})^2.
\end{align*}
The ADADELTA default settings are $\xi =  10^{-6},v=0.95$ and initialize $E(\Delta_{\lambda_i}^2)^{(0)} := E(g_{\lambda_i}^2)^{(0)}=0$. However, in our experiments we obtained better results with $\xi=10^{-7}$.

A popular stopping criterion for the search algorithm is to stop when the moving average lower bound estimates $\overline{\textrm{LB}}_t ={1}/{m} \sum_{i=t-m+1}^\top \widehat{\mathcal{L}(\blamb^{(i)})} $ do not improve after $k$ consecutive iterations \citep{tran2017variational}. Our article
uses $m=k=200$.

\subsection{Details for the Gaussian VB approach.}

Using the factor-based approximation, we can write $\btheta = \bmu + B\beps_1 + d\odot\beps_2$, with $\beps=(\beps_1^\top,\beps_2^\top)^\top\sim N(\boldsymbol{0},\boldsymbol{I}_{r+p})$. Using the reparameterization trick from ($\ref{eq: gradlb}$) and noting that $\gradlogqvb = -(BB^\top+D^2)^{-1}(\btheta-\bmu)$, the gradient of the lower bound $\gradlb$ is
\begin{align*}
	\nabla_{\mu}\lb &= E_f\left[\nabla_{\theta}\log h(\bmu + B\beps_1 + d\odot\beps_2)+(BB^\top+D^2)^{-1}(B\beps_1 + d\odot\beps_2) \right],\\
	\nabla_{B}\lb &= E_f\left[\nabla_{\theta}\log h(\bmu + B\beps_1 + d\odot\beps_2)\beps_1^\top+(BB^\top+D^2)^{-1}(B\beps_1 + d\odot\beps_2)\beps_1^\top \right],\\
	\nabla_{d}\lb &= E_f\left[\diag \left( \nabla_{\theta}\log h(\bmu + B\beps_1 + d\odot\beps_2)\beps_2^\top+(BB^\top+D^2)^{-1}(B\beps_1 + d\odot\beps_2)\beps_2^\top \right) \right],
\end{align*}
where $h(\btheta) = \prior\like$ and $f$ represents $N(\boldsymbol{0},\boldsymbol{I}_{r+p})$. From this, unbiased estimates of the lower bound gradient can be obtained by sampling from $f$. It is necessary to obtain the inverse of the $p\times p$ matrix $(BB^\top+D^2)$, which is computationally expensive when the dimension of $p$ of $\btheta$ is high. Normally, the number of factors we use should be much less than the dimension of $\btheta$, i.e., $r\ll p$. We can then use the Woodbury formula to compute the inverse using \citep{Petersen2012matrix}
$$ (BB^\top+D^2)^{-1} =  D^{-2} - D^{-2}B(I+B^\top D^{-2}B)^{-1}B^\top D^{-2}.$$
This is computationally more efficient because it only requires finding the inverses of the diagonal matrix $D$  and of $(I+B^\top D^{-2}B)$, which is a much smaller $r\times r$ matrix.

\subsection{Details for the Hybrid Gaussian VB}

The gradient of the lower bound with respect to the variational parameters is
\begin{align*}
	\gradlb &=  E_{\beps}\bigg[ \nabla_{\blamb} u(\beps;\blamb)\nabla_{\btheta_1} \log \bigg(\dfrac{p(\btheta_1,\by)}{q_{\blamb}(\btheta_1)}  \bigg) \bigg]\\
	&= E_{(\beps,\btheta_2)}\bigg[ \nabla_{\blamb} u(\beps;\blamb)\nabla_{\btheta_1}\log  \bigg(\dfrac{p(\btheta_1,\by)p(\btheta_2|\btheta_1,\by)}{q_{\blamb}(\btheta_1)p(\btheta_2|\btheta_1,\by)}  \bigg)  \bigg]\\
	&= E_{(\beps,\btheta_2)}\bigg[ \nabla_{\blamb} u(\beps;\blamb)\nabla_{\btheta_1}\log  \bigg(\dfrac{p(\btheta_1,\btheta_2,\by)}{q_{\blamb}(\btheta_1,\btheta_2)}  \bigg)  \bigg]\\
	&= E_{(\beps,\btheta_2)}\bigg[ \nabla_{\blamb} u(\beps;\blamb) \bigg(\nabla_{\btheta_1}\log p(y|\btheta_1,\btheta_2) + \nabla_{\btheta_1}\log p(\btheta_1,\btheta_2) \\
	&\quad\quad\qquad -\nabla_{\btheta_1}\log q_{\blamb}(\btheta_1) - \nabla_{\btheta_1}\log p(\btheta_2|\btheta_1,\by) \bigg)  \bigg].
\end{align*}

Appendix~\ref{appendix:gradients-GVB-Hierarchical-LBA} gives the gradients $\nabla_{\btheta_1} \log p(\by|\btheta_1,\btheta_2),  \nabla_{\btheta_1}\log p(\btheta_1,\btheta_2), \nabla_{\btheta_1}\log q_{\blamb}(\btheta_1)$ and $\nabla_{\btheta_1}\log p(\btheta_2|\btheta_1,\by)$. We note that, because
\[E_{(\beps,\btheta_2)}\bigg[  \nabla_{\btheta_1}\log p(\btheta_2|\btheta_1,\by) \bigg]=E_{\beps}\bigg[\int \nabla_{\btheta_1}p(\btheta_2|\btheta_1,\by) d\btheta_2\bigg]=0,\]
we can remove the term $\nabla_{\btheta_1}\log p(\btheta_2|\btheta_1,\by)$ from the calculation of $\gradlb$. However, this term also plays the role of a control variate and is useful in reducing the variance of the gradient estimate in finite sample sizes (recall we use $N=10$). We therefore include this term in all computations reported in the paper.

\section{\bf Deriving the Gradients in the Gaussian VB for approximating the Hierarchical LBA Models}\label{appendix:gradients-GVB-Hierarchical-LBA}

For the hierarchical LBA model, the joint density of the data $\by$ and model parameters $\btheta = (\balph_1^\top,\dots,\balph_J^\top,\bmualph^\top,\vech(\bSigalph)^\top,\ba)^\top$ is
\begin{align*}
	p(\by,\btheta) &= \like\prior \\
	&=\prod\limits_{j=1}^J \textrm{LBA}(\by_j|\balph_j)N(\balph_j|\bmualph,\bSigalph)N(\bmualph|\boldsymbol{\boldsymbol{0}},\boldsymbol{I})\textrm{IW}(\bSigalph|\nu,\bPsi)\times \prod\limits_{d=1}^{\Da}\textrm{IG}(a_d|1/2,1).
\end{align*}

As mentioned previously, in order to use Gaussian VB, it is necessary to transform the parameters so that all the elements have support on the full real line. The working parameters are $$\tilde{\btheta} = (\balph_1^\top,\dots,\balph_J^\top,\bmualph^\top,\vech(\bCstar)^\top,\log \ba^\top)^\top,$$
where $\log \ba := (\log a_1,\dots,\log a_{\Da})^\top$.
In order to approximate $p(\bthetatilde|y)$ using the Gaussian VB method, it is necessary to have the gradient $\nabla_{\bthetatilde}\log p(\by,\bthetatilde)$ or equivalently, $\nabla_{\bthetatilde} \log p(\by|\bthetatilde)$ and $\nabla_{\bthetatilde} \log p(\bthetatilde)$.

\subsubsection{Computing ${\nabla_{\bthetatilde} \log p(\by|\tilde{\btheta})}$}

Clearly, $\nabla_{\bmualph} \log p(\by|\bthetatilde) = \boldsymbol{0}$, $\nabla_{\vech(\bCstar)} \log p(\by|\bthetatilde)  = \boldsymbol{0}$ and $ \nabla_{\log\ba} \log p(\by|\bthetatilde)  = \boldsymbol{0}$ since $p(\by|\bthetatilde) =\prod\limits_{j=1}^J \textrm{LBA}(\by_j|\balph_j)$ does not depend on the group-level parameters.
\begin{align*}
	\nabla_{\balph_j} \loglike {}&= \dfrac{\partial}{\partial \balph_j} \log \textrm{LBA}(\by_j|\balph_j)= \sum\limits_{i=1}^{n_j} \dfrac{\partial}{\partial \balph_j} \log \textrm{LBA}(y_{ji}|\balph_j)\\ 
	&=\sum\limits_{i=1}^{n_j} \dfrac{ \left( \dfrac{\partial \bz_j}{\partial \balph_j^\top}\right)^\top \dfrac{\partial}{\partial \bz_j} \textrm{LBA}(y_{ji}|\bz_j)}{\textrm{LBA}(y_{ji}|\bz_j)},
\end{align*}
$\dfrac{\partial}{\partial \bz_j} \textrm{LBA}(y_{ji}|\bz_j) = \dfrac{\partial f_c(t)}{\partial \bz_j}(1 - F_{k\neq c}(t)) - \dfrac{\partial F_{k\neq c}(t)}{\partial \bz_j}f_c(t)$


\vspace*{0.5cm}
\hspace*{-0.5cm}The partial derivatives of $f_c(t)$ with respect to $\bz$\footnote{For simplicity, we omit the subscript $j$.} are\\

$\dfrac{\partial}{\partial b}f_c(t) = \dfrac{1}{A}\bigg[  -v^c\xone + s\xtwo + v^c\zone - s\ztwo \bigg]\dfrac{\partial \wone}{\partial b};$\\

$\dfrac{\partial}{\partial A}f_c(t) = -\dfrac{1}{A}\bigg[  f_c(t) - v^c\xone + s\xtwo\bigg]\dfrac{\partial \wone}{\partial A};$\\

$\dfrac{\partial}{\partial v^c}f_c(t) = \dfrac{1}{A}\bigg[  - \xzero +\zzero\bigg] + \dfrac{1}{A}\bigg[  - v^c\xone+s\xtwo+v^c\zone$
\hspace*{2.2cm}	$-s\ztwo\bigg]\dfrac{\partial \wone}{\partial v^c};$\\

$\dfrac{\partial}{\partial \tau}f_c(t) = \dfrac{1}{A}\bigg[   - v^c\xone +s\xtwo \bigg]\left( \dfrac{\partial \wone}{\partial \tau} -\dfrac{\partial \wtwo}{\partial \tau} \right) + \dfrac{1}{A}\bigg[    v^c\zone-s\ztwo\bigg]\dfrac{\partial \wone}{\partial \tau}.$\\

\vspace*{0.5cm}
\hspace*{-0.5cm}The partial derivatives of $F_c(t)$ with respect to $\bz$\footnote{For simplicity, we omit the subscript $j$} are:\\
$\begin{array}{ll}
	
	\dfrac{\partial}{\partial b}F_c(t) &= \bigg[\dfrac{1}{A}\xzero +\uone\xone\dfrac{\partial \wone}{\partial b} \bigg]\\
	&+\dfrac{1}{\wtwo} \bigg[\xtwo-\ztwo \bigg]\dfrac{\partial \wone}{\partial b}+ \bigg[ -\dfrac{1}{A}\zzero - \utwo\zone \dfrac{\partial \wone}{\partial b} \bigg];\\
	
	\dfrac{\partial}{\partial A}F_c(t) &= \dfrac{1}{A}\bigg[-\xzero +\xone(b-A-(t-\tau)v^c)\dfrac{\partial \wtwo}{\partial A} \bigg]\\
	&+\dfrac{1}{A^2}\bigg[ (b-(t-\tau)v^c)\zzero-(b-A-(t-\tau)v^c)\xzero \bigg]\\
	&+ \dfrac{\xtwo\dfrac{\partial \wtwo}{\partial A}\wtwo - \dfrac{\partial \wtwo}{\partial A}\xone }{\wtwo^2}+\zone\dfrac{\partial \wtwo}{\partial A};\\
	
	\dfrac{\partial}{\partial v^c}F_c(t) &= -\dfrac{t-\tau}{A}\xzero+\uone\xone\dfrac{\partial \wone}{\partial v^c}\\
	&+\dfrac{t-\tau}{A}\zzero - \zone\utwo\dfrac{\partial \wone}{\partial v^c}\\
	&+\dfrac{1}{\wtwo}\left( \xtwo-\ztwo\right)\dfrac{\partial \wone}{\partial v^c};\\
	
	\dfrac{\partial}{\partial \tau}F_c(t) &=\dfrac{v^c}{\tau}\xzero+\uone\xone\left( \dfrac{\partial \wone}{\partial \tau} -\dfrac{\partial \wtwo}{\partial \tau} \right) \\
	&-\dfrac{v^c}{\tau}\zzero-\utwo\zone\dfrac{\partial \wone}{\partial \tau} \\
	&+\bigg[\xtwo\wtwo\left( \dfrac{\partial \wone}{\partial \tau} -\dfrac{\partial \wtwo}{\partial \tau} \right)-\xone\dfrac{\partial \wtwo}{\partial \tau} \bigg]/(\wtwo^2)\\
	&-\bigg[\ztwo\wtwo\dfrac{\partial \wone}{\partial \tau}-\zone\dfrac{\partial \wtwo}{\partial \tau} \bigg]/(\wtwo^2).
\end{array}$

\subsubsection*{Computing ${\nabla_{\bthetatilde} \logpriortilde}$}
To get the prior for the transformed parameters $\bthetatilde$, multiply the prior density by the Jacobians:
\begin{align*}
	\priortilde = &p(\balph_{1:J}|\bmualph,\bSigalph)\times p(\bmualph|\bmu,\bSig)\times p(\vech(\bCstar)|\log \ba)\times p(\log \ba)\\
	= &\prod\limits_{j=1}^J N(\balph_{j}|\bmualph,\bSigalph)\times N(\bmualph|\bmu,\bSig)\times \textrm{IW}(\bSigalph|\nu,\bPsi)\times \left|J_{\bSigalph\rightarrow \vech(\bCstar)}\right|\times\cdots \\
	&\times \prod\limits_{d=1}^D \textrm{IG}(a_d|\alpha_d,\beta_d)\times \left|J_{\ba\rightarrow \log \ba} \right|,
\end{align*}
with the prior hyperparameters $\bmu = \boldsymbol{0}$, $\bSig = \boldsymbol{I}_{\Da}$,
$\nu = \nua + \Da - 1$, $\bPsi = 2\nua\diag(1/a_1,\dots,1/a_{\Da})$, $\alpha_d = \dfrac{1}{\mathcal{A}_d^2}$ and $\beta_d = \dfrac{1}{2}$. The Jacobian terms are:
\begin{itemize}
	\item $\left|J_{\bSigalph\rightarrow \vech(\bCstar)}\right| = 2^{\Da}\prod\limits_{d=1}^{\Da}C_{d,d}^{\Da - d + 2},$ with $C_{d,d}$ is an element in posision $(d,d)$ of matrix $\bC$,\\
	and
	\item $\left|J_{\ba\rightarrow \log \ba} \right| =\det \left( \diag(a_1,\dots,a_{\Da})\right) = a_1\times\dots\times a_{\Da}$.
	$\nabla_{\balph_j} \logpriortilde = \nabla_{\balph_j} \log N(\balph_j|\bmualph,\bSigalph) =-\bSigalphinv(\balph_j-\bmualph); $\\
	
	\begin{flalign*}
		\nabla_{\bmualph} \logpriortilde  = &\sum\limits_{j=1}^J \nabla_{\bmualph} \log N(\balph_j|\bmualph,\bSigalph) + \nabla_{\bmualph} \log N(\bmualph|\bmu,\bSig)\\
		= &\sum\limits_{j=1}^J\bSigalphinv(\balph_j-\bmualph) - \bSig^{-1}(\bmualph-\bmu);
	\end{flalign*}
\end{itemize}

\begin{align*}
	&\nabla_{\vech(\bCstar)} \logpriortilde  = \dfrac{\partial}{\partial \vech(\bCstar)} \left[ \sum\limits_{j=1}^J\log N(\balph_j|\bmualph,\bSigalph) + \log \textrm{IW}(\bSigalph|\nu,\bPsi) + \log \left|J_{\bSigalph\rightarrow \vech(\bCstar)}\right|  \right]\\
	=&-\dfrac{J+\nu+p+1}{2}\dfrac{\partial}{\partial \vech(\bCstar)} \log |\bSigalph|-\dfrac{1}{2}\dfrac{\partial}{\partial \vech(\bCstar)}\sum\limits_{j=1}^J(\balph_j-\bmualph)^\top\bSigalphinv(\balph_j-\bmualph)-\cdots\\
	& - \dfrac{1}{2} \dfrac{\partial}{\partial\vech(\bCstar)}\textrm{trace}(\boldsymbol{\Psi}\bSigalphinv) +\dfrac{\partial}{\partial \vech(\bCstar)}\log \left(2^{\Da}\prod\limits_{d=1}^{\Da}C_{d,d}^{\Da - d + 2} \right) \\
	=&-(J+\nu+p+1)\vech(\boldsymbol{I}_{\Da}) - \dfrac{1}{2}\vech(\boldsymbol{\overline{C}}) - \dfrac{1}{2}\vech(\boldsymbol{\overline{\overline{C}}}) + \vech\left(\diag(\Da+1,\Da,\dots,2) \right),
\end{align*}
where $\boldsymbol{\overline{C}}$ and $\boldsymbol{\overline{\overline{C}}}$ are matrices whose elements are
\begin{center}
	$(\boldsymbol{\overline{C}})_{ij} = \left\{ \begin{array}{lr}
		\boldsymbol{M}_{ij}& \textrm{ if } i\neq j\\
		\boldsymbol{M}_{ii}\times (\bC)_{ii} &\textrm{ if } i=j\end{array} \right.$ and  $(\boldsymbol{\overline{\overline{C}}})_{ij} = \left\{ \begin{array}{lr}
		\boldsymbol{H}_{ij}& \textrm{ if } i\neq j\\
		\boldsymbol{H}_{ii}\times (\bC)_{ii} &\textrm{ if } i=j\end{array} \right.,$\\
\end{center}
where
$\boldsymbol{M} = -2\bC^{-T}\bC^{-1}\boldsymbol{\Psi}\bC^{-T}$
and
$\boldsymbol{H} = -2\bC^{-T}\bC^{-1}\sum\limits_{j=1}^J(\balph_j-\bmu)(\balph_j-\bmu)^\top\bC^{-T}.$

\begin{align*}
	&\nabla_{\log a_d}\logpriortilde = \dfrac{\partial}{\partial\log a_d} \bigg[ \log \textrm{IW}(\bSigalph|\nu,\bPsi) + \log \textrm{IG}(a_d|\alpha_d,\beta_d) + \log\left|J_{\ba\rightarrow \log \ba} \right| \bigg]\\
	&= \dfrac{\partial}{\partial\log a_d} \bigg[ \dfrac{\nu}{2}\log\det\left(2\nua\diag(\dfrac{1}{a_1},\dots,\dfrac{1}{a_{\Da}})
	\right) -\dfrac{1}{2}\tr\left( 2\nua\diag(\dfrac{1}{a_1},\dots,\dfrac{1}{a_{\Da}})\bSigalphinv \right) -\cdots\\ &-(\alpha_d+1)\log a_d -\dfrac{\beta_d}{a_d} + \log a_d \bigg]\\
	&=-\dfrac{\nu}{2} - \dfrac{1}{\mathcal{A}^2_d} + \dfrac{1}{2a_d} +\nu_{\alpha}\dfrac{(\bSigalphinv)_{dd}}{a_d}, \textrm{ for } d=1,\dots,D_{\alpha}
\end{align*}

\section{\bf Deriving the Gradients in the Hybrid Gaussian VB for approximating the Hierarchical LBA Models}\label{appendix:gradients-hybrid-GVB}

Recall the set of working parameters is  $\tilde{\btheta}=(\balph_{1:J},\bmualph,\log \ba,\bSigalph)$ which can be partitioned into $\btheta_1 = (\balph_{1:J},\bmualph,\log \ba)$ and $\btheta_2=\bSigalph .$
The data-parameter joint density is
\begin{align*}
	p(\by,\balph_{1:J},\bmualph,\bSigalph,\log \ba) =& \prod\limits_{j=1}^J \textrm{LBA}(\by_j|\balph_j)N(\balph_j|\bmualph,\bSigalph)N(\bmualph|\boldsymbol{\boldsymbol{0}},\boldsymbol{I})\textrm{IW}(\bSigalph|\nu,\bPsi)\\
	&\times \prod\limits_{d=1}^{\Da}\textrm{IG}(a_d|1/2,1)\left| J_{a_d\rightarrow\log a_d}\right|,
\end{align*}
where $J_{a_d\rightarrow\log a_d} = a_d$ is the Jacobian of the transformation.
\begin{lemma}\label{lemma: full-conditional-density}
	For models with random effects given by (\ref{eq:p-y-given-alpha-J})-- (\ref{eq:prior-model-parameters}),   $p(\bSigalph|\balph_{1:J},\bmualph,\ba,\by)$ is the density of $\textrm{IW}(\bSigalph|\nu,\bPsi')$,
	with $\nu = 2\Da +J + 1$ and $\bPsi' = \sum_{j=1}^J (\balph_j - \bmualph)(\balph_j - \bmualph)^\top + 4\diag\left( {1}/{a_1},\dots,{1}/{a_{\Da}}\right) $.
	\begin{proof}
		\begin{align*}
			&p(\bSigalph|\balph_{1:J},\bmualph,\ba,\by)\propto p(\balph_{1:J},\bmualph,\bSigalph,\ba,\by)\propto \prod\limits_{j=1}^J p(\balph_j|\bmualph,\bSigalph)p(\bSigalph|\ba)\\
			&=\prod\limits_{j=1}^J N(\balph_j|\bmualph,\bSigalph)
			\textrm{IW} \left(\bSigalph|\Da +1,\bPsi\right)\qquad \qquad (\textrm{where } \bPsi=4\textrm{diag}(1/a_1,\dots,1/a_{\Da}))\\
			&\propto |\bSigalph|^{-J/2} \exp\left(-\dfrac{1}{2}\sum\limits_{j=1}^J (\balph_j-\bmualph)^\top\bSigalphinv (\balph_j-\bmualph)  \right) |\bSigalph|^{-(2\Da+2)/2}\exp\left(-\dfrac{1}{2}\tr(\bPsi\bSigalphinv) \right)\\
			&\propto |\bSigalph|^{-(2\Da+J+2)/2}\exp\left\{-\dfrac{1}{2}\tr\left( \left(\sum\limits_{j=1}^J (\balph_j-\bmualph)^\top(\balph_j-\bmualph) + \bPsi \right) \bSigalphinv \right) \right\}. 
		\end{align*}
		It is now straightforward to see 
		that $p(\bSigalph|\balph_{1:J},\bmualph,\ba,\by)$ is the density of the Inverse Wishart distribution with the degrees of freedom $\nu = 2\Da + J + 1$ and the scale matrix $\bPsi' = \sum\limits_{j=1}^J (\balph_j-\bmualph)^\top(\balph_j-\bmualph) + 4\textrm{diag}(1/a_1,\dots,1/a_{\Da}).$
	\end{proof}
\end{lemma}
See Appendix \ref{appendix:gradients-GVB-Hierarchical-LBA} for $\nabla_{\btheta_1} \log p(\by|\btheta_1,\btheta_2)$. For the other terms, we first note that
$$p(\btheta_1,\bSigalph) = \prod\limits_{j=1}^J N(\balph_j|\bmualph,\bSigalph)N(\bmualph|\boldsymbol{0},\boldsymbol{I})\textrm{IW}(\bSigalph|\nu,\bPsi) \prod\limits_{d=1}^{\Da}\textrm{IG}(a_d|\alpha_d,\beta_d)\left| J_{a_d\rightarrow\log a_d}\right|;$$
hence,
\begin{align*}
	\log p(\btheta_1,\bSigalph) &= \sum\limits_{j=1}^J \log N(\balph_j|\bmualph,\bSigalph)+ \log N(\bmualph|\boldsymbol{0},\boldsymbol{I})+ \log\textrm{IW}(\bSigalph|\nu,\bPsi)\\
	&+ \sum\limits_{d=1}^{\Da}\Big(\log \textrm{IG}(a_d|\alpha_d,\beta_d) + \log a_d\Big)\\
	&= -\dfrac{1}{2}\sum\limits_{j=1}^J(\balph_j-\bmualph)^\top\bSigalphinv(\balph_j-\bmualph) - \dfrac{1}{2}\bmualph\bmualph^\top +\dfrac{\nu}{2}\logdet(\bPsi)\\
	&-\dfrac{1}{2}\tr (\bPsi\bSigalphinv) +\sum\limits_{d=1}^{\Da}\left\{-(\alpha_d+1)\log a_d - \dfrac{\beta_d}{a_d} + \log a_d\right\}+\text{constant}.
\end{align*}
It follows that
$$\dfrac{\partial}{\partial \balph_j}\log p(\btheta_1,\bSigalph) = -\bSigalphinv(\balph_j-\bmualph),$$
$$\dfrac{\partial}{\partial \bmualph}\log p(\btheta_1,\bSigalph) = \sum\limits_{j=1}^J\bSigalphinv(\balph_j-\bmualph) - \bmualph,$$
and
$$\dfrac{\partial}{\partial \log a_d}\log p(\btheta_1,\bSigalph) = -\dfrac{\nu}{2} - \alpha_d + \dfrac{\beta_d}{a_d} +\nu_{\alpha}\dfrac{(\bSigalphinv)_{dd}}{a_d}.$$

\section{\bf Estimating the posterior predictive densities for the Hierarchical LBA model}\label{appendix:posterior-predictive-densities}
Recall the parameters are $\btheta = (\balph_1^\top,\dots,\balph_J^\top,\bmualph^\top,\vech(\bSigalph)^\top,\ba)^\top$.
\begin{enumerate}
	\item Suppose there are    $\btheta_{MCMC}^{(s)},s=1,\dots,S$ MCMC draws
	from the posterior $\post$;  in the paper $S=9,999$.
	\item Given $\btheta_{MCMC}^{(s)}$, simulate a sample for subject $j$: $\tilde{y}_{js}:=(\textrm{rt}_{js},\textrm{re}_{js})\sim \textrm{LBA}(y|\balph_j^{(s)})$.
	\item Estimate the posterior predictive densities based on the simulated samples $\tilde{\by}_1=(\tilde{y}_{11},\dots, \tilde{y}_{1S}),\dots,\tilde{\by}_J=(\tilde{y}_{J1},\dots, \tilde{y}_{JS})$. In particular:
	\begin{itemize}
		\item The posterior predictive density for the  response time when the decision is correct for subject $j$ under the accuracy condition is estimated based only on the response times corresponding to correct responses performed under accuracy condition. Similar approach can be used to obtain the posterior predictive density for the response time when the decision is correct for subject $j$ under neutral and speed conditions.
		
		\item The posterior predictive density for the response time when the decision is incorrect for accuracy, neutral, and speed conditions can be generated similarly.
	\end{itemize}
\end{enumerate}
The VB posterior predictive densities are estimated similarly, except that  here 
we simulate $\btheta_{VB}^{(s)}\sim \qvb,s=1,\dots,S,$ instead of using the MCMC draws $\btheta_{MCMC}^{(s)}$.

\section{\bf The $K-$fold CVVB applied for hierarchical LBA models}\label{appendix:K-fold CVVB}

\textbf{Input:} A set of LBA models $\{ \mathcal{M}_m\}_{m=1}^M$.\\
\textbf{Output:} The models are ranked based on their predictive power which is measured by $\elpdcvvb$.
\begin{enumerate}
	\item The data is randomly split into $K$ folds. For subject $j$, this is done 
	by  splitting their observations into $K$ disjoint subsets of approximately equal length;
	$$\by_j = \by_{j}^{(1)}\cup\by_{j}^{(2)}\cup\dots\cup\by_{j}^{(K)},\quad j=1,\dots,J. $$
	Denote by $I_j^k$ and $I_j^{-k}$ the set of indices of the observations of subject $j$ that are in fold $k$ and are not in fold $k$, respectively; i.e., the observations in fold $k$ belonging to subject $j$ are $\by_{j}^{(k)} = \left\{y_{ji}|i\in I_j^k  \right\}.$
	Thus,  fold $k$ consists of
	$$\by^{(k)} = \bigcup\limits_{j=1}^J \by_j^{(k)},\quad k=1,\dots,K. $$
	
	\item For each model $\mathcal{M}_m\, (m=1,\dots,M):$
	\begin{itemize}
		\item Leave fold $k$ out, approximate the leave-$k$th-fold-out posterior $\pi(\bthetatilde|\by^{(-k)})$. Denote the VB approximation by  $q_{\blamb^{(k)}}(\bthetatilde)$.
		\item Estimate the leave-$k$th-fold-out posterior predictive density
		$$ p(\by^{(k)}|\by^{(-k)}) \approx  \dfrac{1}{S}\sum\limits_{s=1}^S p(\by^{(k)}|\bthetatilde_{VB}^{(s)}), $$
		\textrm{ where } $$\bthetatilde_{VB}^{(s)} \sim q_{\blamb}(\bthetatilde), s= 1,\dots,S, \textrm{ and } p(\by^{(k)}|\bthetatilde_{VB}^{(s)}) = \prod\limits_{j=1}^J\prod\limits_{i\in I_j^{k}}\textrm{LBA} (y_{ij}| \balph_j^{(s)}).$$
		
		\item The computed K-fold-cross-validation estimate for ELPD is
		$$ \elpdcvvb = \dfrac{1}{K}\sum\limits_{k=1}^K \log \left(\dfrac{1}{S}\sum\limits_{s=1}^S p(\by^{(k)}|\bthetatilde_{VB}^{(s)}) \right).$$
	\end{itemize}
	\item Models are ranked according to the computed K-fold-cross-validation estimate. The model with largest $\elpdcvvb$\ is ranked 1, followed by the second best model which is ranked 2, etc.
\end{enumerate}

\end{document}

%% file: Notations.tex
\def\b1{\boldsymbol{1}}
\def\ba{\boldsymbol{a}}

\def\by{\boldsymbol{y}}

\def\bz{\boldsymbol{z}}

\DeclareMathOperator{\diag}{diag}
\DeclareMathOperator{\logdet}{logdet}
\DeclareMathOperator{\tr}{trace}

\DeclareMathOperator{\vect}{vec}
\DeclareMathOperator{\vech}{vech}

\def\bPsi{\boldsymbol{\Psi}}

\def\sig2{\sigma^2}

\newcommand{\btheta}{\boldsymbol{\theta}}
\newcommand{\bthetatilde}{\boldsymbol{\tilde{\theta}}}

\newcommand{\blamb}{\boldsymbol{\lambda}}

\newcommand{\brho}{\boldsymbol{\rho}}

\newcommand{\nua}{\nu_{\alpha}}

\newcommand{\balph}{\boldsymbol{\alpha}}
\newcommand{\bmualph}{\boldsymbol{\mu_{\alpha}}}
\newcommand{\bmu}{\boldsymbol{\mu}}

\newcommand{\bb}{\boldsymbol{b}}
\newcommand{\bA}{\boldsymbol{A}}
\newcommand{\bv}{\boldsymbol{v}}
\newcommand{\bt}{\boldsymbol{\tau}}

\newcommand{\Da}{D_{\alpha}}
\newcommand{\beps}{\boldsymbol{\epsilon}}

\newcommand{\bSigalph}{\boldsymbol{\Sigma_{\alpha}}}
\newcommand{\bSigalphinv}{\boldsymbol{\Sigma_{\alpha}^{-1}}}
\newcommand{\bSig}{\boldsymbol{\Sigma}}

\newcommand{\bC}{\boldsymbol{C_{\alpha}}}
\newcommand{\bCstar}{\boldsymbol{C^*_{\alpha}}}

\newcommand{\lb}{\mathcal{L}(\boldsymbol{\lambda})}

\newcommand{\lbest}{ \widehat{\mathcal{L}(\blamb)} }
\newcommand{\gradlb}{\nabla_{\boldsymbol{\lambda}}\mathcal{L}(\boldsymbol{\lambda})}

\newcommand{\gradlbest}{\widehat{\nabla_{\boldsymbol{\lambda}}\mathcal{L}(\boldsymbol{\lambda})}}

\newcommand{\mvn}{N_p(\boldsymbol{\mu},\boldsymbol{\Sigma})}

\newcommand{\mvntheta}{N_p(\boldsymbol{\theta}|\boldsymbol{\mu},\boldsymbol{\Sigma})}

\newcommand{\like}{p(\boldsymbol{y}|\boldsymbol{\theta})}
\newcommand{\loglike}{\log p(\boldsymbol{y}|\boldsymbol{\theta})}

\newcommand{\prior}{p(\boldsymbol{\theta})}
\newcommand{\priortilde}{p(\boldsymbol{\tilde{\theta}})}

\newcommand{\logpriortilde}{\log p(\boldsymbol{\tilde{\theta}})}

\newcommand{\joint}{p(\boldsymbol{y},\boldsymbol{\theta})}

\newcommand{\post}{p(\boldsymbol{\theta}|\boldsymbol{y})}

\newcommand{\qvb}{q_{\boldsymbol{\lambda}}(\boldsymbol{\theta})}

\newcommand{\gradlogqvb}{\nabla_{\btheta} \log q_{\boldsymbol{\lambda}}(\boldsymbol{\theta})}

\newcommand{\logmarg}{\log \widehat{p(\by)}}
\newcommand{\elpdcv}{\widehat{\textrm{ELPD}}_{\textrm{K-CV}}}
\newcommand{\elpdcvvb}{\widehat{\textrm{ELPD}}_{\textrm{K-CVVB}}}

\newcommand{\uone}{\dfrac{b-A-(t-\tau)v^c}{A}}
\newcommand{\utwo}{\dfrac{b-(t-\tau)v^c}{A}}

\newcommand{\xzero}{\Phi(\overline{\omega}_1-\overline{\omega}_2)}
\newcommand{\xone}{\phi(\overline{\omega}_1-\overline{\omega}_2)}
\newcommand{\xtwo}{\phi'(\overline{\omega}_1-\overline{\omega}_2)}

\newcommand{\zzero}{\Phi(\overline{\omega}_1)}
\newcommand{\zone}{\phi(\overline{\omega}_1)}
\newcommand{\ztwo}{\phi'(\overline{\omega}_1)}

\newcommand{\wone}{\overline{\omega}_1}
\newcommand{\wtwo}{\overline{\omega}_2}